\documentclass[a4paper,10pt]{article}

\usepackage[textwidth=18cm, textheight=26cm]{geometry}
\usepackage[utf8]{inputenc}
\usepackage[english]{babel}
\usepackage[T1]{fontenc}
\usepackage{graphicx}
\usepackage{lmodern}
\usepackage[hidelinks]{hyperref}
\usepackage{amsmath}
\usepackage{amssymb}
\usepackage{bm}
\usepackage{siunitx}
\usepackage{cleveref}
\usepackage{booktabs}
\usepackage[affil-it]{authblk}
\usepackage{titlesec}

\makeatletter
\patchcmd{\@maketitle}{\LARGE \@title}{\fontsize{14}{16.8}\selectfont\@title}{}{}
\makeatother

\titleformat{\section}{\normalfont\fontsize{10}{12}\bfseries}{\thesection}{0.5em}{}
\titleformat{\subsection}{\normalfont\fontsize{10}{12}\itshape}{\thesubsection}{0.5em}{}
\titleformat{\subsubsection}{\normalfont\fontsize{10}{12}\itshape}{\thesubsubsection}{0.5em}{}

\hypersetup{pdfauthor=author}

\def\etal.{et\penalty50\ al.}

\begin{document}

\title{A variational data assimilation approach for sparse velocity reference data in coarse RANS simulations through a corrective forcing term}

\author[,a]{Oliver Brenner\footnote{Corresponding author: \href{mailto:brennero@ethz.ch}{\texttt{brennero@ethz.ch}}}}
\author[a,b]{Justin Plogmann}
\author[,a]{Pasha Piroozmand\footnote{Present address: IBU Institut für Bau und Umwelt, Ostschweizer Fachhochschule, Oberseestrasse 10, Rapperswil, CH-8640, Switzerland}}
\author[a]{Patrick Jenny}

\affil[a]{Institute of Fluid Dynamics, ETH Zurich, Sonneggstrasse 3, Zürich, CH-8092 Switzerland}
\affil[b]{Chemical Energy Carriers and Vehicle Systems Laboratory, Swiss Federal Laboratories for Materials Science and Technology (Empa), Überlandstrasse~129, Dübendorf, CH-8600 Switzerland}

\date{\fontsize{11}{13.2}\selectfont April 30, 2024}


\maketitle


\noindent\hrulefill

\section*{Abstract}

The Reynolds-averaged Navier--Stokes (RANS) equations provide a computationally efficient method for solving fluid flow problems in engineering applications.
However, the use of closure models to represent turbulence effects can reduce their accuracy.
To address this issue, recent research has explored data-driven techniques such as data assimilation and machine learning.

An efficient variational data assimilation (DA) approach is presented to enhance steady-state eddy viscosity based RANS simulations.
To account for model deficiencies, a corrective force term is introduced in the momentum equation.
In the case of only velocity reference data, this term can be represented by a potential field and is divergence-free.
The DA implementation relies on the discrete adjoint method and approximations for efficient gradient evaluation.

The implementation is based on a two-dimensional coupled RANS solver in \emph{OpenFOAM}, which is extended to allow the computation of the adjoint velocity and pressure as well as the adjoint gradient.
A gradient-based optimizer is used to minimize the difference between the simulation results and the reference data.
To evaluate this approach, it is compared with alternative data assimilation methods for canonical stationary two-dimensional turbulent flow problems.
For the data assimilation, sparsely distributed reference data from averaged high-fidelity simulation results are used.

The results suggest that the proposed method achieves the optimization goal more efficiently compared to applying data assimilation for obtaining the eddy viscosity, or a field modifying the eddy viscosity, directly.
The method works well for different reference data configurations and runs efficiently by leveraging coarse meshes.

\noindent\hrulefill


\section{Introduction}

There are numerous relevant fluid flow problems in science and engineering, ranging from the smallest scales, e.\,g. in biological flows, to atmospheric flows at a planetary scale.
Efficient and accurate analysis of fluid flow problems is critical in many engineering applications, including power generation with gas turbines or coal combustion, renewable power generation such as hydro and wind, and energy storage such as pumped-storage hydropower.
Efficient propulsion is essential in transportation, utilizing jet engines, propellers, and internal combustion engines, alongside streamlined aircraft, ships, and land vehicles to achieve low drag and minimize noise (e.\,g.~\cite{spalart16,stopford02,costes12,lain22,mani23}).

Analysis of these flows often relies on simulation, since experiments might not be feasible for technical, financial, or safety reasons.
Compared to experimental investigations, flow simulations have further advantages such as their flexibility and their speed, but they never will exactly represent reality.
Many of the aforementioned technical applications involve turbulent flows, which are challenging to simulate because of the large number of spatial and temporal scales involved.
Direct numerical simulation or high fidelity methods such as large eddy simulation (LES) are prohibitively expensive for many industrially relevant problems, while Reynolds-averaged Navier--Stokes (RANS) based simulations are widely used.
RANS computations only provide (ensemble) averaged results and rely upon more modeling and approximations than high-fidelity methods, but they are significantly less expensive.

The RANS approach in particular is prone to a number of errors related to averaging and the closure models, which often rely on the Boussinesq approximation that introduces the eddy viscosity.
Two equation models for the eddy viscosity, such as the $k$-$\varepsilon$ and $k$-$\omega$ models (cf.~\cite{wilcox06} or \cite{pope00}) or later the SST model~\cite{menter94}, or the one equation Spalart Allmaras model~\cite{spalart92}, are widely used, despite some limitations like, e.\,g., issues with strong streamline curvatures in the $k$-$\varepsilon$ model \cite{craft96}.

More recent developments have focused on data driven models or machine learning to close the RANS equations (cf.~\cite{duraisamy18}), and the related field of uncertainty quantification (UQ) (e.\,g. \cite{xiao17,xiao19a}).
The goal of Physics-informed machine learning (PIML) (e.\,g.~\cite{karniadakis21}) is to integrate physical knowledge into machine learning methods, such as neural networks or kernel-based methods.
Wang~\etal.~\cite{wang17a} utilized a PIML approach to correct modeled Reynolds-stresses in RANS equations, demonstrating its effectiveness even in cases of extrapolation, where training and prediction occur in different geometries.

Physics-informed neural networks (PINN) were introduced by Raissi~\etal.~\cite{raissi18} with the aim of replacing numerical partial differential equation (PDE) solvers for physical problems with neural networks.
These networks take spatial coordinates (and time) as inputs and provide the quantities of interest, e.\,g. velocity and pressure, at these locations (and times) as output.
What makes PINNs distinct from other neural network based methods is the training, which directly involves the PDEs of the physical problem, e.\,g. the Navier--Stokes equations, with initial and boundary conditions through automatic differentiation.
Sliwinski and Rigas~\cite{sliwinski23} extended this to the RANS equations with a solenoidal forcing and sparsely distributed reference data for pressure and velocity.
They present very good results for the mean flow reconstruction around a circular cylinder.

Another approach to combine data with simulations is data assimilation (DA), which allows to tune simulation parameters such that the results better coincide with the reference.
Data assimilation involves solving a inverse problems, since the model input (the parameters) is sought from the model output; the simulation model is then referred to as the forward problem.
Any kind of model parameter, such as closure model constants, and initial or boundary conditions, can be adjusted in this manner.
Here, it is not a matter of finding a general closure model, but rather of tuning a specific simulation setup with reference data, e.\,g. from a corresponding measurement.
This is for example used in numerical weather prediction (cf. \cite{kwon18}) or in the analysis of biological flows, e.\,g. \cite{koltukluoglu18,epp20,averweg22}.

DA methods can broadly be divided into variational and statistical methods.
Variational DA aims to reduce a cost function that measures the difference between model output and reference by tuning the model parameters, while statistical DA tries to reduce the variability of the result~\cite{asch16}.
Nudging is a less sophisticated DA approach that was introduced for weather forecasting~\cite{asch16}.
For large parameter spaces, statistical DA methods such as the Kalman-filter, are prohibitively expensive.
A suitable method for such instances is the adjoint method (e.\,g.~\cite{bradley13}), a variational approach with a computational cost that does not scale with the number of parameters.
Yang~\etal.~\cite{yang15} demonstrated that ensemble-based variational data assimilation, combining variational and statistical approaches, offers advantages over conventional variational methods within the context of a shallow water model

Data assimilation with the adjoint method is based on a minimization problem with the objective of reducing a cost function that measures the difference between model output and reference data, as well as regularization contributions.
This minimization is based on the cost function gradient with respect to the parameters, which is provided by the adjoint method at a cost that is comparable to the cost of solving the forward problem.
However, as the adjoint method is a model intrusive approach, deriving the expression for the adjoint gradient can be laborious and is problem specific.
The discretized expressions for the adjoint problem and gradient can be obtained by either initially deriving them analytically and then discretizing (e.\,g. \cite{hafez22}), the so called \emph{continuous} method, or by directly deriving them from the discretized forward problem (e.\,g. \cite{fleischli21,dilgen18}) as in the \emph{discrete} method.

The application of adjoint data assimilation to close the RANS equations or to reconstruct flow fields from (sparse) experimental data has been subject to a number of studies presented in the literature.
An important question is the choice of data assimilation parameters, which was very recently studied by Cato~\etal.~\cite{cato23}.
They investigated six different parameter choices for data assimilation with the discrete adjoint method in RANS simulations and compared their performance for three different flow setups.
Another choice is the degree of modelling a data assimilation method is relying on.
Foures~\etal.~\cite{foures14}, for example, have introduced a continuous adjoint data assimilation approach to reconstruct the divergence of the Reynolds stress tensor field in the steady RANS equations from experimental reference data.
They use this as physics based interpolation to recover full field information from sparse experimental mean flow measurements, without relying on the Boussinesq hypothesis or RANS closure models.
Li~\etal.~\cite{li22} have extended this approach for cases with only limited numbers of wall pressure measurements.
More recently Patel~\etal.~\cite{patel24} have compared the adjoint approach by Foures~\etal.~\cite{foures14} to a PINN approach by Sliwinski and Rigas~\cite{sliwinski23} for mean flow reconstruction.
As an extension, they avoid direct reconstruction of the divergence of Reynolds stresses in the RANS equations, opting instead for a forcing term that corrects deviations in Reynolds stresses calculated by an eddy viscosity model, following the approach introduced by Franceschini~\etal.~\cite{franceschini20a}.
In their study, Franceschini~\etal.~\cite{franceschini20a} compared this forcing term to an alternative source term in the eddy viscosity transport equation of the Spalart-Almaras model for reconstructing mean flow fields in various high-Reynolds number flows.
For discrete adjoint based DA, Brenner~\etal.~\cite{brenner22} have introduced further simplifications for eddy viscosity models based on \emph{OpenFOAM} operators.
Other than the choice of the data assimilation parameters, the optimal placement of reference data (cf.~\cite{papadimitriou15,mons17}) and the regularization of the parameter field (cf.~\cite{epp22,piroozmand23}) are important questions.
Leoni~\etal.~\cite{leoni20} used nudging based DA to study the required type and distribution of sparse reference data to reconstruct flow features in homogeneous and isotropic turbulence.

In this work, we introduce a \emph{discrete} adjoint method based data assimilation approach for RANS problems that relies on existing closure models.
The objective is to establish a data assimilation framework that enhances a low-cost simulation, allowing for precise full field simulation results at reasonable computational cost.
This is achieved by relying on the RANS equations and coarse meshes for the forward problem, on a single scalar parameter field for tuning, and an efficient data assimilation implementation in \emph{OpenFOAM} based on the discrete adjoint method.
In particular, a parameterized forcing term is introduced in the Reynolds stress closure term to correct for eddy viscosity model errors, thus not limiting the approach by the eddy viscosity assumption.
This is similar to the approach by Patel~\etal.~\cite{patel24} that use a divergence free force term with the \emph{continuous} adjoint method.
The performance of this approach is assessed by comparing it to alternative data assimilation approaches that act directly on the eddy viscosity.
All these approaches aim at improving a particular \emph{simulation} setup and not the underlying closure models.

This paper is organized as follows.
\Cref{sec:methods} introduces the proposed data assimilation approach with all the necessary components, as well as two alternative approaches.
One of the alternative approaches is based on modifying the eddy viscosity with a multiplicative field.
A recap of this method is presented in \cref{sec:app_alpha} together with a more detailed discussion of the modifications made over the original approach.
In \cref{sec:results} a flow over a backward facing step is used to compare the proposed data assimilation method with its alternatives.
This is followed by results with the corrective forcing approach at flows over a periodic hill and a bump setup.
Additional results for the periodic hill setup with the modified eddy viscosity and the corrective forcing term are presented in \cref{sec:app_ph_alpha,sec:app_ph_psi}, respectively.
Finally, \cref{sec:conclusion} discusses conclusions and potential areas for future research.


\section{Methods}
\label{sec:methods}

\subsection{Simulation of turbulent flows}

\subsubsection{Reynolds-averaged Navier--Stokes equations}

Reynolds-averaging separates a quantity $\theta$ into its average $\bar{\theta}$ and fluctuation $\theta'$.
Applying this to the continuity and momentum equations of a steady incompressible Newtonian fluid results in
\begin{equation}
    \label{eq:rans_continuity}
    \frac{\partial \bar{u}_{j}}{\partial x_{j}}
    =
    0
\end{equation}
and
\begin{equation}
    \label{eq:rans_momentum}
    \frac{\partial \bar{u}_{i} \bar{u}_{j}}{\partial x_{j}}
    +
    \frac{\partial}{\partial x_{i}}
    \left[
        \frac{\bar{p}}{\rho}
    \right]
    -
    \frac{\partial}{\partial x_{j}}
    \left[
        2 \nu \bar{S}_{ij}
    \right]
    +
    \frac{\partial \overline{u'_{i} u'_{j}} }{\partial x_{j}}
    =
    0
\end{equation}
with the unclosed Reynolds stresses $\overline{u'_{i} u'_{j}}$ and the averaged rate-of-strain tensor
\begin{equation}
    \bar{S}_{ij}
    =
    \frac{1}{2}
    \left(
        \frac{\partial \bar{u}_{i}}{\partial x_{j}}
        +
        \frac{\partial \bar{u}_{j}}{\partial x_{i}}
    \right)
    \, .
\end{equation}
Alternatively, \cref{eq:rans_continuity,eq:rans_momentum} can be formulated as a system of momentum and pressure equations.
The latter is obtained by applying a divergence operator on the momentum equation, followed by a simplification step based on the continuity equation.

\subsubsection{Turbulence modeling}

To close the RANS \cref{eq:rans_momentum}, the Boussinesq hypothesis
\begin{equation}
    \label{eq:boussinesq}
    \overline{u_{i}'u_{j}'}
    =
    \frac{2}{3}k\delta_{ij}
    -
    2\nu_{t}\bar{S}_{ij}
\end{equation}
with the turbulent kinetic energy $k$ and the Kronecker delta $\delta_{ij}$ is used.
This introduces the eddy viscosity $\nu_{t}$ that relates the Reynolds stress tensor to the averaged rate-of-strain tensor $\bar{S}_{ij}$.
For the steady momentum equation, this yields
\begin{equation}
\label{eq:rans_momentum_nut}
    \frac{\partial \bar{u}_{i} \bar{u}_{j}}{\partial x_{j}}
    +
    \frac{\partial p^{*}}{\partial x_{i}}
    -
    \frac{\partial}{\partial x_{j}}
    \left[
        2 \left(\nu + \nu_{t}\right) \bar{S}_{ij}
    \right]
    =
    0
    \, ,
\end{equation}
with the modified pressure
\begin{equation}
    \label{eq:rans_pressure_mod}
    p^{*}
    =
    \frac{\bar{p}}{\rho}
    +
    \frac{2}{3} k
    \, .
\end{equation}
The corresponding pressure equation is derived from this momentum equation and the continuity \cref{eq:rans_continuity} as
\begin{equation}
\label{eq:rans_pressure_nut}
    R_{p^{*}}
    =
    \frac{\partial \bar{u}_{i}}{\partial x_{j}}
    \frac{\partial \bar{u}_{j}}{\partial x_{i}}
    +
    \frac{\partial^{2} p^{*}}{\partial x_{i} \partial x_{i}}
    -
    \frac{\partial}{\partial x_{i}}
    \left[
    \frac{\partial}{\partial x_{j}}
        \left[
            2 \nu^{\mathrm{eff}} \bar{S}_{ij}
        \right]
    \right]
    =
    0
\end{equation}
with the effective viscosity
\begin{equation}
    \nu^{\mathrm{eff}}
    =
    \nu
    +
    \nu_{t}
    \, .
\end{equation}

The $k$-$\varepsilon$ model is used in this work to obtain the eddy viscosity based on the solution of transport equations for the turbulent kinetic energy $k$ and the turbulent rate of dissipation $\varepsilon$.
In particular, the default \emph{coupledKEpsilon} model of \emph{foam-extend-5.0} is used.
This implementation is based on a coupled solution of these two extra equations.

\subsection{Data assimilation with a corrective forcing term}
\label{subsec:methods_psi_da}

Closure models based on the Boussinesq hypothesis (cf. \cref{eq:boussinesq}) are often successful in engineering applications, but they still suffer from shortcomings.
In this work, we introduce a forcing term $F_{i}$ to correct for discrepancies in the divergence of the modeled Reynolds stress tensor as
\begin{equation}
    \frac{\partial \overline{u'_{i} u'_{j}}}{\partial x_{j}}
    =
    \frac{\partial}{\partial x_{j}}
    \left(
        \frac{2}{3}k\delta_{ij}
        -
        2\nu_{t}\bar{S}_{ij}
    \right)
    -
    F_{i}
    =
    \frac{2}{3}
    \frac{\partial k}{\partial x_{i}}
    -
    \frac{\partial}{\partial x_{j}}
    \left(
        2\nu_{t}\bar{S}_{ij}
    \right)
    -
    F_{i}
\end{equation}
and subject it to a Stokes--Helmholtz decomposition, i.\,e.,
\begin{equation}
    \label{eq:helmholtz}
    F_{i}
    =
    -\frac{\partial \phi}{\partial x_{i}}
    +
    \epsilon_{ijk} \frac{\partial \psi_{k}}{\partial x_{j}}
\end{equation}
with a scalar potential $\phi$, a vector potential $\psi_{k}$, and the Levi-Civita symbol $\epsilon$.
In this work, only velocity reference data is assimilated, which only affects the $\psi_{k}$ parameter, whereas the assimilation of pressure reference data would affect the $\phi$ parameter.
This naturally leads to the forcing term being divergence free.

For the residual of the steady momentum equation this yields
\begin{equation}
    \label{eq:psi_rans_momentum}
    R_{\bar{u}_{i}}
    =
    \frac{\partial \bar{u}_{i} \bar{u}_{j}}{\partial x_{j}}
    +
    \frac{\partial p^{\dag}}{\partial x_{i}}
    -
    \frac{\partial}{\partial x_{j}}
    \left[
        2\left(\nu + \nu_{t}\right)\bar{S}_{ij}
    \right]
    -
    \epsilon_{ijk} \frac{\partial \psi_{k}}{\partial x_{j}}
    =
    0
    \, ,
\end{equation}
where the averaged pressure $\bar{p}$, the isotropic part of the Reynolds stress tensor, and the scalar potential $\phi$ (cf. \cref{eq:helmholtz}) are absorbed in the modified pressure $p^{\dag}$ as
\begin{equation}
    \label{eq:psi_rans_pressure_mod}
    p^{\dag}
    =
    \frac{\bar{p}}{\rho}
    +
    \frac{2}{3} k
    +
    \phi
    \, .
\end{equation}
This pressure definition effectively eliminates parameter $\phi$ from the equations, only leaving parameter $\psi_{k}$ for data assimilation.
Furthermore, due to the divergence-free property of the $\psi_{k}$ term, the resulting pressure equation is identical to the one in the basic RANS approach in \cref{eq:rans_pressure_nut}, i.\,e., parameter $\psi_{k}$ is not present in the pressure equation.

Compared to the approach by Foures~\etal.~\cite{foures14} only $\psi_{k}$ appears explicitly in the momentum equations, and the discrete adjoint approach is used.
Also, the Reynolds stress tensor still is primarily based on the eddy viscosity model and is only corrected by data assimilation where the model is not sufficient.
Further, data assimilation is performed for $\psi_{k}$, such that no additional equations need to be solved for $\phi$ and $\psi_{k}$, as in~\cite{li22}.
For the optimization, the parameter field $\psi_{k}$ is thus initialized with a uniform zero value, corresponding to no correction of the eddy viscosity model.
In the case of two-dimensional RANS only the $\psi_{z}$ component is relevant, i.\,e., a scalar field is sufficient to influence both velocity components, which reduces complexity over directly tuning two components of a force $F_{i}$.
This is different than in the work by Patel~\etal.~\cite{patel24}, who also use a divergence-free forcing term.
Further differences to their work include the regularization and the parameter mapping.

\subsection{Alternative data assimilation parameters}

Two further parameter options for data assimilation are presented below.
They serve to use as benchmark for the corrective forcing term approach.
The first choice is a slightly modified version of the method by Brenner~\etal.~\cite{brenner22} which introduces a spatially variable parameter field to locally tune the frozen eddy viscosity field obtained from the $k$-$\varepsilon$ closure model.
The second choice relies on tuning the eddy viscosity directly.

\subsubsection{Modified eddy viscosity}

More concretely, an initial solution of the RANS \cref{eq:rans_continuity} is computed, which serves as baseline solution with corresponding eddy viscosity $\nu_{t}$.
A spatially variable parameter field $\alpha$ is then introduced in the RANS equation as
\begin{equation}
    \label{eq:alpha_rans_momentum}
    R_{\bar{u}_{i}}
    =
    \frac{\partial \bar{u}_{i} \bar{u}_{j}}{\partial x_{j}}
    +
    \frac{\partial p^{\ddag}}{\partial x_{i}}
    -
    \frac{\partial}{\partial x_{j}}
    \left[
        2 \left(
            \nu
            +
            \alpha \nu_{t}^{b}
        \right)
        \bar{S}_{ij}
    \right]
    =
    0
\end{equation}
with the frozen \emph{baseline} eddy viscosity $\nu_{t}^{b}$ and the modified pressure $p^{\ddag}$.
In this case, the pressure equation will deviate from the baseline expression in \cref{eq:rans_pressure_nut} with parameter $\alpha$ in the viscous term.
To simplify notation $\bar{p}$ is used for any of the above modified pressures, i.\,e., $p^{*}$ in \cref{eq:rans_pressure_mod,eq:alpha_rans_momentum}, $p^{\dag}$ in \cref{eq:psi_rans_pressure_mod}, and $p^{\ddag}$ in \cref{eq:alpha_rans_momentum}, in the remainder of the manuscript.

For the optimization, the parameter field $\alpha$ is uniformly initialized with a value of unity across the entire domain.
A more detailed description of the method and the modifications made over the previous work by Brenner~\etal.~\cite{brenner22} are presented in \cref{sec:app_alpha}.

\subsubsection{Tuned eddy viscosity}

A more direct approach is to not use a closure model and directly tune the eddy viscosity $\nu_{t}$ in \cref{eq:rans_momentum_nut,eq:rans_pressure_nut}.
To obtain a converged initial flow solution, a suitable initial value for this eddy viscosity field needs to be selected.
In the absence of any prior knowledge a positive uniform field is assumed, which corresponds to an increase in viscosity or a reduction of the Reynolds number.
It is thus important to choose an initial value for the eddy viscosity field that is as small as possible, but sufficiently large to obtain a converged solution, i.\,e., $0 < \nu_{t}^{\left(0\right)} \ll 1$.

\subsection{Forward problem formulation}

To obtain the numerical solutions of \cref{eq:rans_continuity,eq:rans_momentum}, the momentum equations for $\bar{u}_{i}$ and the pressure equation for $\bar{p}$ are solved.
Consequently, the solution vector is defined as
\begin{equation}
    U
    =
    \left[
        \bar{u}, \,
        \bar{p}
    \right]^{T}
    =
    \left[
        \bar{u}_{x}, \,
        \bar{u}_{y}, \,
        \bar{u}_{z}, \,
        \bar{p}
    \right]^{T}
    \, .
\end{equation}
In each iteration the forward system of equations is linearized and solved in a coupled manner as
\begin{equation}
    \label{eq:coupled_residual}
    R\left(\xi, U\right)
    =
    \begin{bmatrix}
        R_{\bar{u}}\left(\xi, U\right)\\
        R_{\bar{p}}\left(\xi, U\right)
    \end{bmatrix}
    =
    \begin{bmatrix}
        \mathbf{A}_{\bar{u}\bar{u}} & \mathbf{A}_{\bar{u}\bar{p}}\\
        \mathbf{A}_{\bar{p}\bar{u}} & \mathbf{A}_{\bar{p}\bar{p}}
    \end{bmatrix}
    \begin{bmatrix}
        \bar{u}\\
        \bar{p}
    \end{bmatrix}
    -
    \begin{bmatrix}
        b_{\bar{u}}\\
        b_{\bar{p}}
    \end{bmatrix}
    =
    \mathbf{A}_{U}
    U
    -
    b_{U}
    =
    0
    \, ,
\end{equation}
where $\xi$ represents a generic parameter used for data assimilation.
The system matrix of the coupled linear system of equations $\mathbf{A}_{U}$ is composed of sub-matrices that describe the implicit contributions of the velocity ($\mathbf{A}_{\bar{u}\bar{u}}$) and pressure ($\mathbf{A}_{\bar{u}\bar{p}}$) in the momentum equations, and the contributions of the velocity ($\mathbf{A}_{\bar{p}\bar{u}}$) and pressure ($\mathbf{A}_{\bar{p}\bar{p}}$) in the pressure equation, respectively.
The discretization, i.\,e., construction of the linear system of equations, and the solution of the same equations is based on the finite volume code \emph{OpenFOAM}.

\subsection{Inverse problem formulation}

A scalar cost functional $f$ is introduced as a measure for the agreement between model output and reference data.
In this work, the cost function consists of two contributions, namely a regularization function $f_{\xi}$ that only depends on the parameter field $\xi$, and a discrepancy contribution $f_{U}$ that is a measure for the difference between the forward solution $U$ and the reference, respectively, i.\,e.,
\begin{equation}
    \label{eq:cost_function}
    f\left(\xi, U\right)
    =
    f_{\xi}\left(\xi\right)
    +
    f_{U}\left(U\right)
    .
\end{equation}

Based on this cost function, the data assimilation procedure is formulated as an optimization problem
\begin{subequations}
    \begin{alignat}{2}
        & \!\min_{\xi}      & \quad & f\left(\xi, U\right)\\
        & \text{subject to} &       & R\left(\xi, U\right) = 0 \, ,
    \end{alignat}
\end{subequations}
with the objective of reducing the discrepancy between the forward problem solution and the reference data by tuning the parameter field $\xi$.
This is an inverse problem as the input parameter $\xi$ for the forward problem is sought based on the results $U$ obtained from the forward problem.

\subsubsection{Cost function and regularization}

The discrepancy part of the cost function measures agreement of the forward problem solution $U$ with the reference data $U^{\mathrm{ref}}$.
In the presented application, only averaged velocity data is assimilated, i.\,e.,
\begin{equation}
    \label{eq:discrepancy}
    f_{U}\left(U\right)
    =
    \frac{
        1
    }{
        V^{\mathrm{ref}}
    }
    \sum\limits_{j\in\mathcal{R}} \left[
        \sum\limits_{k\in\left\{x,y\right\}}
        \left(
            \bar{u}_{k,j}
            -
            \bar{u}_{k,j}^{\mathrm{ref}}
        \right)^{2}
        V_{j}
    \right]
    \, ,
\end{equation}
where $\mathcal{R}$ is the set of reference cells $j$, $V_{j}$ the volume of cell $j$, and
\begin{equation}
    V^{\mathrm{ref}}
    =
    \sum\limits_{j\in\mathcal{R}}
    V_{j}
\end{equation}
the total volume of all reference cells.

A regularization function is introduced to reduce ambiguity of the inverse problem.
Since the flow setups investigated in this work are only two-dimensional and in the $x$-$y$ plane, only the $z$-component of vector potential $\psi_{k}$ has a non-zero value.
Thus, regularization only is applied to the $\psi_{z}$ component of the parameter field.
This is done without loss of generality, i.\,e., the other components would be subject to the same regularization in the general three-dimensional case.
In particular, following the method by Brenner~\etal.~\cite{brenner22}, a function of the form
\begin{equation}
    \label{eq:regularization}
    f_{\xi}\left(\xi\right)
    =
    w^{\mathrm{reg}}
    \sum\limits_{i\in\Omega} \left[
        \frac{1}{\left|\mathcal{B}_{i}\right|}
        \sum\limits_{k\in\mathcal{B}_{i}} \left(
            \xi_{i}
            -
            \xi_{k}
        \right)^{2}
    \right]
\end{equation}
with regularization weight $w^{\mathrm{reg}}$ is used to guide the optimization toward smooth parameter fields.
Here, index $i$ loops over all cells in the simulation domain $\Omega$, and index $k$ loops over $\mathcal{B}_{i}$, the set of neighboring cells of cell $i$, with $\left|\mathcal{B}_{i}\right|$ denoting the number of neighboring cells of cell $i$.

\subsubsection{Test function}

Apart from the reference data utilized in the cost function and the consequent optimization procedure, a second set of reference data is used.
This collection, called the \emph{test} dataset, of reference values that are not used for data assimilation, but instead serve for the assessment of the test function $f^{\mathrm{test}}(U)$.
The form of this function is equivalent to the discrepancy part of the cost function, i.\,e., of $f_{U}(U)$ in \cref{eq:discrepancy}.

Conceptually, this is similar to the idea of using training, validation, and test data sets in the training of machine learning models.
The regularization in \cref{eq:regularization} features a hyper parameter $w^{\mathrm{reg}}$ that requires tuning and that introduces a trade-off.
Larger values for $w^{\mathrm{reg}}$ promote smooth parameter fields, which results in more accurate solution results between the sparse reference data locations, while smaller values result in lower values of the discrepancy part of the cost function $f_{U}$.

The test function helps to balance this trade-off by measuring the quality of the optimized solution in-between reference locations.
A good regularization weight is found, when both cost and the test functions are reduced during the optimization.

\subsection{Inverse problem solution}

\subsubsection{Discrete adjoint method}

To derive an expression for the cost function gradient, a Lagrangian
\begin{equation}
    \label{eq:lagrangian}
    \mathcal{L}\left(\xi, U\right)
    =
    f\left(\xi, U\right)
    -
    \lambda^{T} R\left(\xi, U\right)
\end{equation}
with Lagrangian multiplier
\begin{equation}
    \label{eq:lambda}
    \lambda
    =
    \left[
        \lambda_{\bar{u}}, \,
        \lambda_{\bar{p}}
    \right]^{T}
    =
    \left[
        \lambda_{\bar{u}_{x}}, \,
        \lambda_{\bar{u}_{y}}, \,
        \lambda_{\bar{u}_{z}}, \,
        \lambda_{\bar{p}}
    \right]^{T}
\end{equation}
is introduced.
The corresponding gradient with respect to the parameters $\xi$ is derived as
\begin{equation}
    \label{eq:adjoint_gradient}
    \frac{\mathrm{d} f}{\mathrm{d} \xi}
    =
    \frac{\partial f_{\xi}}{\partial \xi}
    -
    \lambda^{T} \frac{\partial R}{\partial \xi}
    \, .
\end{equation}
For a more detailed derivation and description, the reader is referred to \cref{sec:app_adjoint} and \cite{brenner22}.

Rearranging the terms and considering that the derivative of the linearized forward problem residual $R$ with respect to the linearized forward problem solution $U$ corresponds to the respective system matrix $\mathbf{A}_{U}$ (cf. \cref{eq:coupled_residual}) yields
\begin{equation}
    \label{eq:coupled_adjoint}
    \mathbf{A}_{U}^{T} \, \lambda
    =
    \frac{\partial f_{U}}{\partial U}^{T}
    \, .
\end{equation}
The right-hand side of this equation is analytically derived from the cost function and thus comes at low computational cost.
Solving this system of linear equations for $\lambda$ has an associated computational cost that is comparable to solving the forward system for $U$.
It is important to note that this cost is independent of the number of parameters $\xi$.

Applied to the RANS forward problem with residual \cref{eq:coupled_residual} and cost function \cref{eq:cost_function} with \cref{eq:discrepancy}, the \emph{coupled} adjoint system of equations, i.\,e. \cref{eq:coupled_adjoint}, reads
\begin{equation}
    \renewcommand\arraystretch{1.2}
    \begin{bmatrix}
        \frac{\partial R_{\bar{u}}}{\partial \bar{u}}^{T} & \frac{\partial R_{\bar{p}}}{\partial \bar{u}}^{T}\\
        \frac{\partial R_{\bar{u}}}{\partial \bar{p}}^{T} & \frac{\partial R_{\bar{p}}}{\partial \bar{p}}^{T}
    \end{bmatrix}
    \begin{bmatrix}
        \lambda_{\bar{u}}\\
        \lambda_{\bar{p}}
    \end{bmatrix}
    =
    \begin{bmatrix}
        \frac{\partial f_{\bar{u}}}{\partial \bar{u}}^{T}\\
        \frac{\partial f_{\bar{p}}}{\partial \bar{p}}^{T}
    \end{bmatrix}
    =
    \begin{bmatrix}
        \frac{\partial f_{\bar{u}}}{\partial \bar{u}}^{T}\\
        0
    \end{bmatrix}
    \, .
\end{equation}

The derivatives $\frac{\partial R}{\partial \xi}$ of the forward problem residual with respect to the parameters are evaluated using the approximate and efficient approach introduced by Brenner~\etal.~\cite{brenner22}.
In particular, the forward problem $R$ is numerically linearized with respect to parameter $\xi$ in \emph{OpenFOAM} as
\begin{equation}
    R
    =
    \mathbf{A}_{\xi} \, \xi
    -
    b_{\xi}
    =
    0
    \, .
\end{equation}
For the derivative with respect to parameter $\xi$, that is needed for the evaluation of the adjoint gradient in \cref{eq:adjoint_gradient}, this yields
\begin{equation}
    \frac{\partial R}{\partial \xi}
    =
    \frac{\partial}{\partial \xi}
    \left[
        \mathbf{A}_{\xi} \, \xi
        -
        b_{\xi}
    \right]
    =
    \mathbf{A}_{\xi}
    \, .
\end{equation}
Combining these numerically approximated terms with \cref{eq:adjoint_gradient} and the solution of \cref{eq:coupled_adjoint} and expanding the terms results in
\begin{equation}
    \label{eq:rans_adjoint_gradient}
    \frac{\mathrm{d} f}{\mathrm{d} \xi}
    =
    \frac{\partial f_{\xi}}{\partial \xi}
    -
    \lambda_{\bar{u}_{x}}^{T} \mathbf{A}_{\bar{u}_{x},\xi}
    -
    \lambda_{\bar{u}_{y}}^{T} \mathbf{A}_{\bar{u}_{y},\xi}
    -
    \lambda_{\bar{p}}^{T} \mathbf{A}_{\bar{p},\xi}
\end{equation}
for the discrete adjoint gradient.
As a simplification, Brenner~\etal.~\cite{brenner22} did not consider the last term in \cref{eq:rans_adjoint_gradient}, i.\,e., the pressure contribution.
While this is justified in their periodic hill case, where this contribution is negligible, it is considered in this work.

In the case of the corrective forcing, the parameter $\psi_{k}$ only appears in a single term in the momentum equation $R_{\bar{u}}$ (cf. \cref{eq:psi_rans_momentum}), i.\,e.,
\begin{equation}
    \label{eq:dRdPsi}
    \frac{\partial R_{\bar{u}}}{\partial \psi}
    =
    \frac{\partial}{\partial \psi}
    \left[
        -\epsilon_{ijk} \frac{\partial \psi_{k}}{\partial x_{j}}
    \right]
    \approx
    -\frac{\partial}{\partial \psi}
    \left[
        \mathbf{A}_{\bar{u},\psi} \psi - b_{\psi}
    \right]
    =
    -\mathbf{A}_{\bar{u},\psi}
    \, ,
\end{equation}
where $\mathbf{A}_{\bar{u},\psi}$ is the system matrix and $b_{\psi}$ the right-hand side vector of the implicit \emph{curl} operator, respectively.
Due to the divergence-free property of the forcing term, there is no contribution of parameter $\psi_{k}$ to the pressure equation (cf. \cref{subsec:methods_psi_da}).
The derivative of the pressure residual with respect to the parameter, i.\,e., $\frac{\partial R_{\bar{u}}}{\partial \psi}$, is thus zero and there is no contribution of the adjoint pressure $\lambda_{\bar{p}}$ to the adjoint gradient.
The implicit curl operator required in \cref{eq:dRdPsi} is not available by default in any version of \emph{OpenFOAM} and was thus specifically implemented and validated for this work (cf. \cref{sec:app_imp_curl}).

For the modified eddy viscosity approach with parameter $\alpha$, the corresponding gradient evaluation yields
\begin{equation}
    \label{eq:dRUdAlpha}
    \frac{\partial R_{\bar{u}}}{\partial \alpha}
    =
    \frac{\partial}{\partial \alpha}
    \left[
        -\frac{\partial}{\partial x_{j}}
        \left(
            \left(
                \nu_{t}
                \bar{S}_{ij}
            \right)
            \alpha
        \right)
    \right]
    \approx
    -\frac{\partial}{\partial \alpha}
    \left[
        \mathbf{A}_{\bar{u},\alpha} \alpha - b_{\alpha}
    \right]
    =
    -\mathbf{A}_{\bar{u},\alpha}
    \, ,
\end{equation}
where matrix $\mathbf{A}_{\bar{u},\alpha}$ is thus given by the system matrix of the implicit \emph{divergence} operator of \emph{OpenFOAM}.
In contrast to the corrective forcing term, the pressure contribution is non-zero here.
In particular, it is derived as
\begin{align}
    \label{eq:dRPdAlpha}
    \nonumber
    \frac{\partial R_{\bar{p}}}{\partial \alpha}
    &=
    \frac{\partial}{\partial \alpha}
    \left[
        -\frac{\partial}{\partial x_{i}}
        \left(
            \frac{\partial}{\partial x_{j}}
            \left(
                \left(
                    \nu_{t}
                    \bar{S}_{ij}
                \right)
                \alpha
            \right)
        \right)
    \right]\\
    &=
    -\frac{\partial}{\partial \alpha}
    \left[
        \underbrace{
            \frac{\partial}{\partial x_{i}}
            \left(
                \left(
                    \nu_{t}
                    \bar{S}_{ji}
                \right)
                \frac{\partial}{\partial x_{j}} \alpha
            \right)
        }_{\text{Laplacian}}
        +
        \underbrace{
            \frac{\partial}{\partial x_{i}}
            \left(
                \left(
                    \nu_{t}
                    \frac{\partial \bar{S}_{ij}}{\partial x_{j}}
                \right)
                \alpha
            \right)
        }_{\text{divergence}}
    \right]
    \approx
    -\frac{\partial}{\partial \alpha}
    \left[
        \mathbf{A}_{\bar{p},\alpha} \alpha - b_{\alpha}
    \right]
    =
    -\mathbf{A}_{\bar{p},\alpha}
    \, .
\end{align}
Matrix $\mathbf{A}_{\bar{p},\alpha}$ is thus built from the implicit \emph{Laplacian} and \emph{divergence} operators in \emph{OpenFOAM}.

In the case where the eddy viscosity is directly adjusted within the data assimilation process, the derivations in \cref{eq:dRUdAlpha,eq:dRPdAlpha} are analogous.

\subsubsection{Parameter map}

Depending on the choice of optimization parameter, the range of parameter values might be limited by physical or other constraints.
In this work, only positive values are allowed for the DA parameters $\alpha$ and $\nu_{t}$ to prevent numerical instabilities, while no such constraint is necessary for parameter $\psi_{z}$.

This can be addressed by introducing a regularization function $f_{\xi}$ that punishes values outside the permitted limits with large values.
However, this approach can not strictly guarantee that these limits are not violated.

Following Brenner~\etal.~\cite{brenner22}, a problem dependent mapping function is thus introduced to neatly integrate these limits into the discrete adjoint framework.
In particular, the \emph{physical} parameter field $\xi$, to which the limits apply, is mapped to the \emph{optimization} parameter field $\zeta\in\mathbb{R}$ as $\xi = \xi\left(\zeta\right)$.
The gradient based parameter updates are thus performed for $\zeta$ and then mapped back to $\xi$.
This update step requires the cost function gradient with respect to the optimization parameter $\zeta$, which is obtained by mapping the adjoint gradient with respect to $\xi$ using the product rule.
Evaluating these forward and inverse parameter maps, as well as the gradient mapping, is trivial and comes at very low computational cost.
It only involves the numerical evaluation of the corresponding expressions which are derived analytically from the expression of the parameter map.

Here, the exponential-linear map proposed by Brenner~\etal.~\cite{brenner22} is applied for DA parameters $\alpha$ and $\nu_{t}$, while no map (i.\,e. a linear map) is applied for $\psi_{z}$.

\subsubsection{Optimization}

The gradient based Demon-Adam optimization algorithm~\cite{chen19} is employed to perform updates of optimization parameter $\zeta$ from iteration step $\left(n\right)$ to $\left(n+1\right)$ as
\begin{equation}
    \zeta_{i}^{\left(n+1\right)}
    =
    \zeta_{i}^{\left(n\right)}
    -
    \Delta\zeta
    \left(
        \left.\frac{\mathrm{d} f}{\mathrm{d}\zeta}\right|^{\left(n\right)}_{i}
    \right)
    \, .
\end{equation}
The parameter update $\Delta\zeta$ is a function of the adjoint gradient value and is determined based on the optimization algorithm.

Convergence of the optimizer can be defined based on the cost function value, the gradient norm, the number of iterations, or other measures.
In this work, the maximum number of iterations is selected as a-priori convergence criterion for better comparisons between the different approaches and setups.

\subsection{Reference data}

Literature reference data available from public online resources was used for assimilation.
Accordingly, the test case setups are replicating the reference case geometries and boundary conditions.
In particular, spanwise and temporally averaged LES or DNS velocity reference data was used.

The \emph{griddata} interpolation function for unstructured data provided by the \emph{SciPy} library for \emph{Python} was used for interpolation of the reference data to the cell center locations of the \emph{OpenFOAM} meshes.
In particular, the nearest neighbor method was employed and the full fields were interpolated.
To select the reference and test data sets from these full fields, two indicator fields are introduced, to mark test and reference cells, respectively.

\subsection{Implementation}

The implementation is based on \emph{foam-extend-5.0}~\cite{fe50}, a version of the open-source CFD framework \emph{OpenFOAM}.
This specific version was selected because it offers coupled solvers, e.\,g. for the RANS equations.
In particular, the \emph{pUCoupledFoam} solver was modified and extended for the different DA parameters in this work.
The extension, which was introduced in \cite{brenner22}, includes a periodic forcing term to drive stationary flows in periodic domains.
Three solvers were implemented, one for each DA parameter.
In the case of parameter $\psi_{k}$, this warranted the implementation of an implicit curl operator (cf. \cref{sec:app_imp_curl}).
These solvers were modified to solve the RANS forward equations and the adjoint equations, as well as to evaluate the cost function and the adjoint gradient.
Considering the two-dimensional equations considered in this work, all solvers are dedicated two-dimensional implementations.
This reduces the size of the linear systems of equations and thus the memory usage as well as the linear solver cost.

\begin{figure}[!ht]
    \centering
    \includegraphics[width=\textwidth]{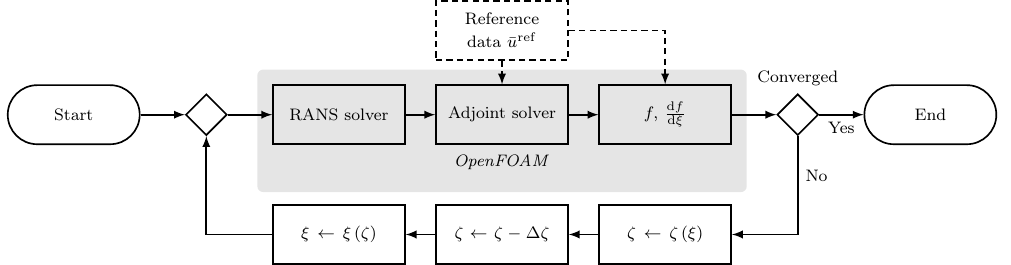}
    \caption{
        Flowchart of the optimization based data assimilation procedure.
        Starting from an initial RANS setup, the parameter field $\xi$ is updated iteratively based on the cost function gradient $\mathrm{d} f/\mathrm{d}\xi$, until convergence is reached.
        An \emph{OpenFOAM} solver is used to evaluate the forward and adjoint problems, as well as the cost function and its gradient (gray shading).
        To enforce bounds for $\xi$, parameter updates are performed for optimization parameter $\zeta$ which is connected to $\xi$ by a mapping function.
        The parameter updates, the parameter mapping, and the optimization are implemented in a \emph{Python} script.
    }
    \label{fig:diagram_rans}
\end{figure}

\Cref{fig:diagram_rans} gives an overview of the complete data assimilation process.
The process is controlled by a \emph{Python} script that interacts with the \emph{OpenFOAM} solvers and performs the optimization loop including the mapping of parameters and gradients.

To create the meshes, the \emph{OpenFOAM-8}~\cite{of8} utility \emph{blockMesh} was used.
This version of \emph{OpenFOAM} provides a more extensive implementation of the utility, allowing more control over the mesh generation process.

\section{Results and discussion}
\label{sec:results}

In this section, results for the three data assimilation parameter choices that were introduced above are presented.
The results for a backward facing step setup with two different reference and test data distributions are compared to discuss the effectiveness of the different methods.
Then, further results for the corrective forcing term method in a periodic hill and in a bump setup are presented.
Further results can be found in \cref{sec:app_ph_psi,sec:app_ph_alpha}.

\subsection{Comparison of data assimilation approaches for a backward facing step setup}

\begin{figure}[!ht]
    \centering
    \includegraphics[width=\textwidth]{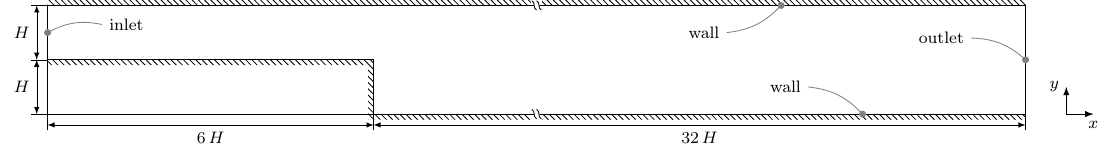}
    \caption{
        The backward facing step domain as presented by Pont-Vílchez~\etal.~\cite{pontvilchez19} with labels for the different boundaries.
        All length scales are normalized by step height $H$.
    }
    \label{fig:bfs_geometry}
\end{figure}

The simulation domain of the backward facing step setup used as a test case, and presented in \cref{fig:bfs_geometry}, is based on the work by Pont-Vílchez~\etal.~\cite{pontvilchez19}.
Time-averaged DNS velocity results of their work serve as reference data here.
The setup features an inlet channel of height $H$, followed by a backward facing step with expansion ratio \num{2}.
Considering the viscosity $\nu$ and the inlet bulk velocity $\bar{u}_{b}$, the Reynolds number is
\begin{equation}
    \mathrm{Re}
    =
    \frac{\bar{u}_{b} H}{\nu}
    =
    \num{6932}
    \, .
\end{equation}
The domain is spatially discretized to obtain a coarse mesh of \num{4800} cells.

The boundary conditions for the boundaries indicated in \cref{fig:bfs_geometry} are based on the reference data case.
All walls feature no-slip boundary conditions for velocity $\bar{u}$, Neumann boundary conditions for pressure $\bar{p}$, and wall models for turbulent kinetic energy $k$ and the turbulence dissipation $\varepsilon$.
The inlet velocity profile is interpolated from the reference data and velocity fluctuations at the inlet are assumed to have a magnitude of \SI{5}{\percent} of the average inlet velocity, and the turbulent length scale is assumed to be \SI{10}{\percent} of the inlet channel height $H$, to compute values for $k$ and $\varepsilon$, respectively.
Zero gradient boundary conditions at the outlet are chosen for all fields except pressure, where a fixed value is chosen for pressure.

\begin{figure}[!ht]
    \centering
    \includegraphics[width=0.49\textwidth]{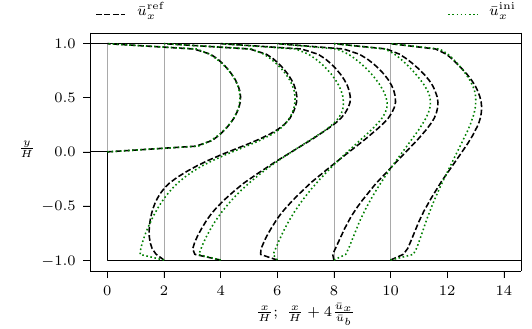}
    \hfill
    \includegraphics[width=0.49\textwidth]{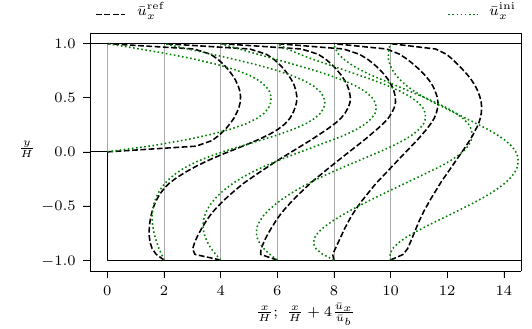}
    \caption{
        Stream wise velocity components of the reference $\bar{u}^{\mathrm{ref}}$, and the baseline RANS results $\bar{u}^{\mathrm{ini}}$ for the backward facing step setup at $\mathrm{Re} = \num{6932}$.
        The left plot shows the baseline RANS solution, while the right plot shows the solution for the initially uniform eddy viscosity $\nu_{t}^{\left(0\right)}$ of the direct eddy viscosity tuning approach.
        All length scales are normalized by the inlet channel or step height $H$ and all velocities by the inlet bulk velocity $\bar{u}_{b}$, respectively.
    }
    \label{fig:bfs_baseline_result}
\end{figure}

The baseline RANS solution with a turbulence model results in velocity profiles $\bar{u}_{x}^{\mathrm{ini}}$ that deviate from the reference $\bar{u}_{x}^{\mathrm{ref}}$, especially in the wake of the step, as depicted in \cref{fig:bfs_baseline_result}.
Much better agreement with the reference is found in the upstream channel, which is attributed to the velocity inlet boundary condition that is interpolated from the reference data.
Further downstream of the step, where its effects fade off, the agreement with the reference improves again.

\begin{figure}[!ht]
    \centering
    \includegraphics[width=0.49\textwidth]{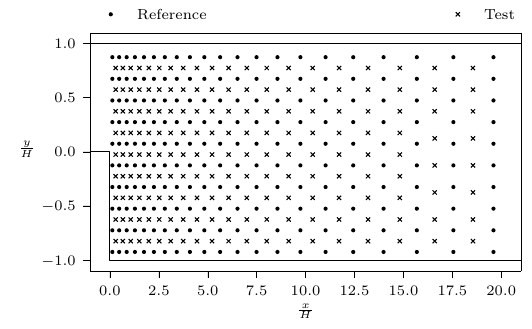}
    \hfill
    \includegraphics[width=0.49\textwidth]{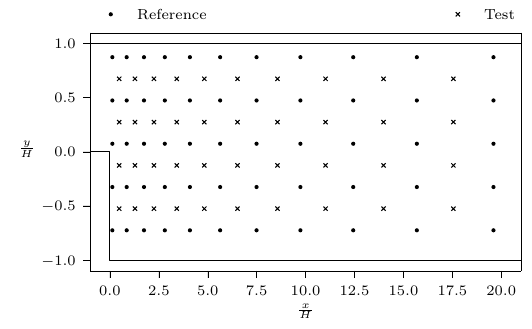}
    \caption{
        Distribution of \emph{reference} (dots) and \emph{test} (crosses) data points for the backward facing step in the \emph{dense} (left) and \emph{sparse} (right) setups.
    }
    \label{fig:bfs_ref_map}
\end{figure}

\Cref{fig:bfs_ref_map} depicts the two reference and test data distributions considered in this section, that is, a \emph{dense} distribution with \num{210} reference and \num{178} test points, and a \emph{sparse} distribution with \num{55} reference and \num{40} test points, respectively.
These reference points are concentrated in the section just downstream of the step, where most of the relevant flow dynamics occurs and where turbulence modelling is most challenging.
The corresponding test points are placed in-between the reference points to best measure the effects of data assimilation and regularization on the flow field where no information is assimilated.

Next, data assimilation results for this backward facing step setup with both \emph{dense} and \emph{sparse} reference distributions are presented for all three approaches.

\subsubsection{Data assimilation for eddy viscosity}

\begin{figure}[!ht]
    \centering
    \includegraphics[width=0.49\textwidth,trim=0 22 0 0,clip]{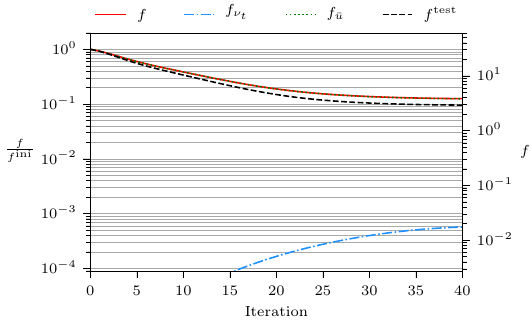}
    \hfill
    \includegraphics[width=0.49\textwidth,trim=0 22 0 0,clip]{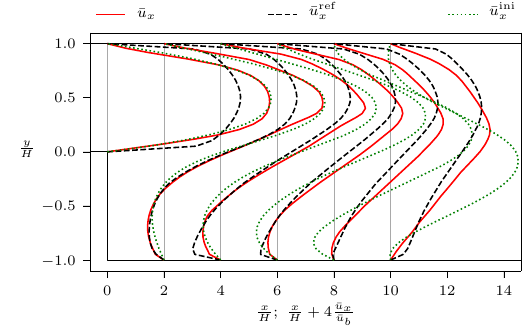}\\
    \includegraphics[width=0.49\textwidth,trim=0 0 0 13,clip]{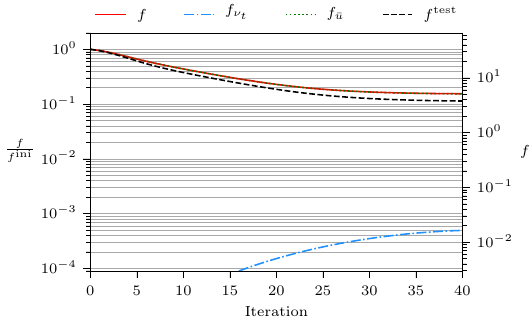}
    \hfill
    \includegraphics[width=0.49\textwidth,trim=0 0 0 13,clip]{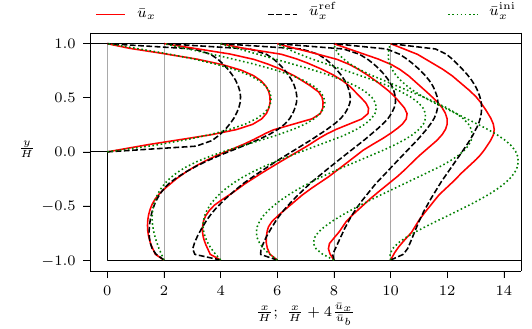}
    \caption{
        Data assimilation results with parameter $\nu_{t}$ for the backward facing step (cf. \cref{fig:bfs_geometry}) with \emph{dense} (top) and \emph{sparse} (bottom) reference data distributions (cf. \cref{fig:bfs_ref_map}).
        The left row shows the cost function components and the test function.
        The left-hand side ordinate shows the cost function component values and test function values normalized by the initial value for the cost function and the test function, respectively.
        The right hand-side ordinate shows the original values of the cost function components.
        In the right row, the stream wise velocity components of the reference $\bar{u}^{\mathrm{ref}}$, the baseline RANS results $\bar{u}^{\mathrm{ini}}$, and the optimized results $\bar{u}$ are shown.
        Note that the cost function $f$ is the sum of the regularization $f_{\nu_{t}}$ and discrepancy $f_{U}$ contributions.
        All length scales are normalized by the inlet channel or step height $H$ and all velocities by the inlet bulk velocity $\bar{u}_{b}$, respectively.
    }
    \label{fig:bfs_nut_results}
\end{figure}

For data assimilation parameter $\nu_{t}$, no RANS turbulence model is solved during the forward solution process.
To initialize the eddy viscosity field, a value of $\nu_{t}^{\left(0\right)}=\SI{5e-02}{\meter\squared\per\second\squared}$ was used, based on the principle of selecting the smallest possible uniform field that still leads to a converged RANS solution.
Compared with the fluid viscosity of $\nu=\SI{2.53e-03}{\meter\squared\per\second\squared}$ (i.\,e. $\nu_{t}^{\left(0\right)}/\nu=\num{19.76}$), this is equivalent to a drastic reduction of the effective Reynolds number to $\mathrm{Re}^{\mathrm{eff}}=\num{333.87}$, which explains the strong deviation of the corresponding velocity profiles $\bar{u}^{\mathrm{ini}}$ from the reference velocity $\bar{u}^{\mathrm{ref}}$ in \cref{fig:bfs_nut_results}.

Throughout this results section the optimization is performed until convergence is reached based on the slope of the cost and test functions.
A learning rate of $\eta=\num{1e-03}$, a number of total iteration steps of $N^{\mathrm{opt}}=\num{40}$, and a regularization weight of $w^{\mathrm{reg}}=\num{1e-3}$ is used in this case.

As seen in the cost function plots of \cref{fig:bfs_nut_results}, the test and cost functions get continuously but slowly reduced for both reference configurations during the optimization.
The velocity profile plots show that both optimized results strongly deviate from the reference.
Further, it can be observed that the regularization works well here, as the test function continuously decreases and smooth optimized velocity profiles are obtained.
The agreement with the reference, however, is very poor.
To conclude, this approach proved to be unsuccessful for data assimilation in the case of this setup.

\subsubsection{Data assimilation for eddy viscosity correction}
\label{subsub:res_bfs_nut}

Next, the performance of the modified eddy viscosity for the backward facing step setup is investigated, again for the two configurations presented in \cref{fig:bfs_ref_map}.
The baseline eddy viscosity field $\nu_{t}^{b}$ results from solving the forward problem with the $k$-$\varepsilon$ model in a precursor step.
Not surprisingly, the resulting initial velocity field $\bar{u}^{\mathrm{ini}}$ is much closer to the reference data than in the previous case, as depicted in \cref{fig:bfs_psi_results}.
The eddy viscosity field is then \emph{frozen}, i.\,e., not updated, when solving the forward problem during the optimization.
An initially uniform parameter field of $\alpha^{\left(0\right)}=\num{1}$ is chosen, and the optimization is performed with a learning rate of $\eta=\num{1e-03}$, a number of total iteration steps of $N^{\mathrm{opt}}=\num{25}$, and a regularization weight of $w^{\mathrm{reg}}=\num{5e-05}$.

\begin{figure}[!ht]
    \centering
    \includegraphics[width=0.49\textwidth,trim=0 22 0 0,clip]{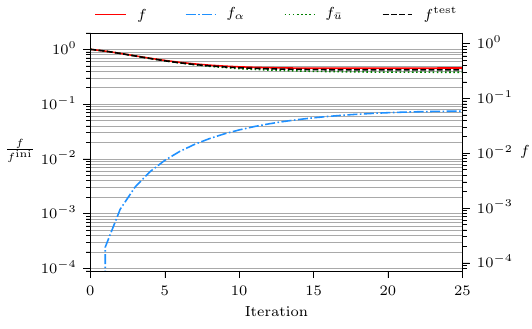}
    \hfill
    \includegraphics[width=0.49\textwidth,trim=0 22 0 0,clip]{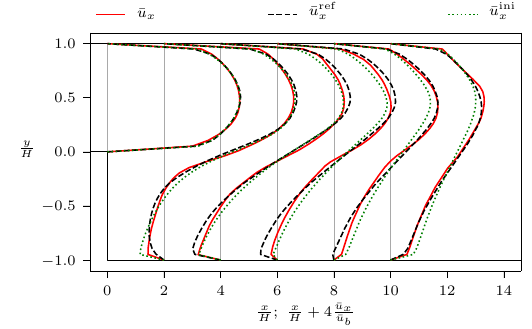}\\
    \includegraphics[width=0.49\textwidth,trim=0 0 0 13,clip]{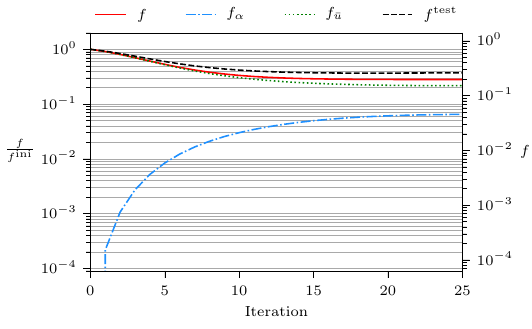}
    \hfill
    \includegraphics[width=0.49\textwidth,trim=0 0 0 13,clip]{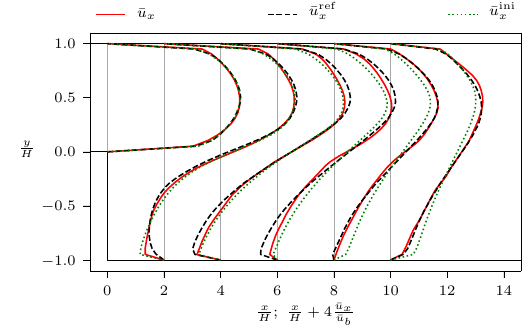}
    \caption{
        Data assimilation results with parameter $\alpha$ for the backward facing step (cf. \cref{fig:bfs_geometry}) with \emph{dense} (top) and \emph{sparse} (bottom) reference data distributions (cf. \cref{fig:bfs_ref_map}).
        Shown are the cost function components and the test function (left), as well as the streamwise velocity components of the reference $\bar{u}^{\mathrm{ref}}$, the baseline RANS results $\bar{u}^{\mathrm{ini}}$, and the optimized results $\bar{u}$ (right).
    }
    \label{fig:bfs_alpha_results}
\end{figure}

The resulting cost and test function plots in \cref{fig:bfs_psi_results} indicate a continuous improvement throughout the domain, but it is evident from the velocity profiles that this is mainly happening downstream of the recirculation zone.
In the recirculation zone, the baseline solution deviates most from the reference, and data assimilation hardly improves the results there, in particular in the case of the \emph{sparse} reference data distribution.

The final results are much closer to the reference, if the eddy viscosity correction factor is tuned instead of the eddy viscosity directly.
Nevertheless, this approach falls short of producing a very close match.
While data assimilation for $\alpha$ was reported to provide good results for a well-resolved periodic hill case, it seems to perform much worse in this case.
Like the eddy viscosity assumption in general, the influence of parameter $\alpha$ is limited to flow regions of large mean shear rates.

\subsubsection{Data assimilation for corrective forcing}

\begin{figure}[!ht]
    \centering
    \includegraphics[width=0.49\textwidth,trim=0 22 0 0,clip]{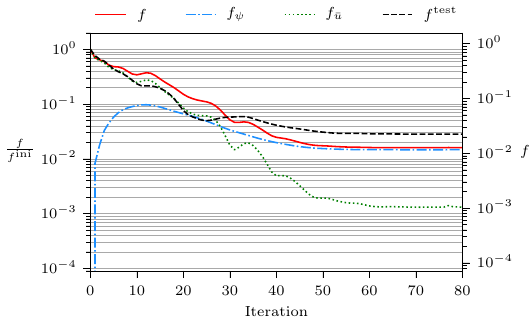}
    \hfill
    \includegraphics[width=0.49\textwidth,trim=0 22 0 0,clip]{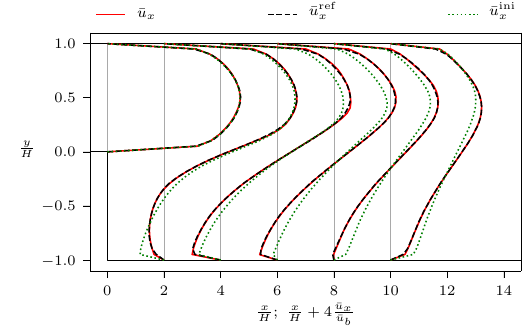}\\
    \includegraphics[width=0.49\textwidth,trim=0 0 0 13,clip]{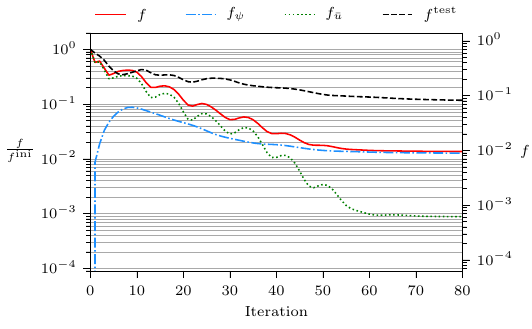}
    \hfill
    \includegraphics[width=0.49\textwidth,trim=0 0 0 13,clip]{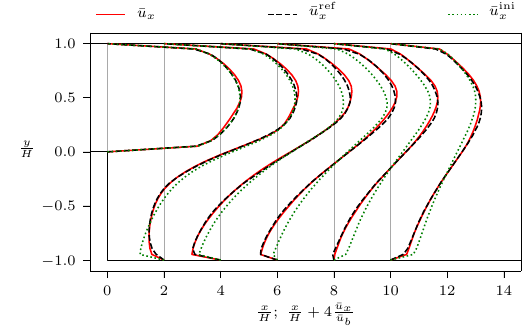}
    \caption{
        Data assimilation results for parameter $\psi_{k}$ with the backward facing step (cf. \cref{fig:bfs_geometry}) with \emph{dense} (top) and \emph{sparse} (bottom) reference data distributions (cf. \cref{fig:bfs_ref_map}).
        Shown are the cost function components and the test function (left), as well as the stream wise velocity components of the reference $\bar{u}^{\mathrm{ref}}$, the baseline RANS results $\bar{u}^{\mathrm{ini}}$, and the optimized results $\bar{u}$ (right).
    }
    \label{fig:bfs_psi_results}
\end{figure}

Finally, the corrective forcing approach is applied to the backward facing step setups.
Starting from an initial uniform parameter field $\psi_{z}^{\left(0\right)}=\num{0}$, the same initial RANS solution as in the previous case (cf. \cref{subsub:res_bfs_nut}) is seen, as depicted in \cref{fig:bfs_psi_results}.
Data assimilation is subsequently performed with a learning rate of $\eta=\num{1e-02}$, a number of total iteration steps of $N^{\mathrm{opt}}=\num{40}$, and a regularization weight of $w^{\mathrm{reg}}=\num{4e-05}$.
Contrary to the other approaches, the modeled eddy viscosity is updated in every optimization step.

The cost and test function plots in \cref{fig:bfs_psi_results} get drastically reduced during the optimization procedure, clearly surpassing the previously demonstrated results.
From the corresponding velocity profile plots it is clear that the \emph{dense} reference data lead to excellent agreement with the reference data, while there still are slight deviations present in the recirculation zone for the \emph{sparse} case.

\subsection{Discussion of the backward facing step results}

This section demonstrates the differences in data assimilation performance between the three approaches.
Directly working on the eddy viscosity that is initialized with a uniform field leads to large deviations in the initial solution, which are not significantly reduced through data assimilation.

Tuning the eddy viscosity obtained from a proven physics-based closure model is more promising.
The initial solution is much closer to the reference and is further improved through data assimilation.
However, for the present setup, the improvement is not satisfactory.
What makes this setup difficult for RANS solvers is the coarse mesh resolution and the sharp edge at the step with the associated recirculation.

The corrective forcing proves to be the most powerful approach.
It combines a good initial condition of the closure model with a source term that can influence the flow beyond the limitations of an eddy viscosity model.
The coarse mesh resolution is no limitation and thus allows for very good data assimilation results at low cost.

\subsection{Further results for data assimilation with corrective forcing}

In this section, results for data assimilation with the corrective forcing approach for two further flow problems are presented, namely the \emph{periodic hill} and the \emph{bump} setups.
As for the backward facing step above, directly tuning $\nu_{t}$ starting with an initially uniform parameter field does not yield meaningful results for any of the two setups.
For the periodic hill setup, correcting the eddy-viscosity with parameter $\alpha$ can produce convincing results, as shown by Brenner~\etal.~\cite{brenner22}.
However, for the coarse mesh used here, the improvements are less convincing, which is further discussed in \cref{sec:app_ph_alpha}.
It will be shown that for the bump setup, the $\alpha$ approach fails to produce accurate results.

\subsubsection{Periodic hill}
\label{subsubsec:res_ph}

\begin{figure}[!ht]
    \centering
    \includegraphics[width=0.49\textwidth]{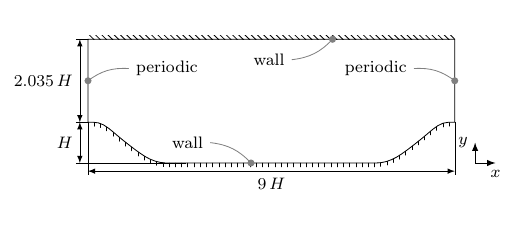}
    \hfill
    \includegraphics[width=0.49\textwidth,trim=0 0 0 10,clip]{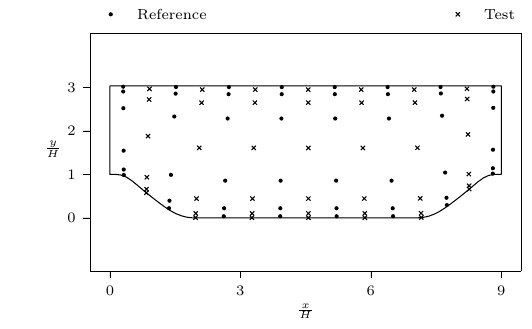}
    \caption{
        Simulation domain of the periodic hill setup with indicated boundary types (left).
        Mean flow is in $x$ direction.
        All length scales are normalized by the hill height $H$.
        Distributions of \emph{reference} (dots) and \emph{test} (crosses) data points for the \emph{sparse} configuration are shown on the right.
    }
    \label{fig:ph_geometry_ref_map}
\end{figure}

The periodic hill geometry with its corresponding \emph{sparse} reference data distribution is depicted in \cref{fig:ph_geometry_ref_map}.
A coarse mesh of \num{74} cells in $x$-direction by \num{60} cells in $y$-direction (total cell count is \num{4440}) is used with \num{48} reference and \num{42} test points that are sparsely distributed in the domain.
Velocity data from LES simulations by Gloerfelt and Cinnella~\cite{gloerfelt19}, averaged in time and in $z$-direction, are considered as reference.
This setup features a flow at a Reynolds number of
\begin{equation}
    \mathrm{Re}
    =
    \frac{\bar{u}_{b}H}{\nu}
    =
    \num{10595}
    \, ,
\end{equation}
based on the bulk velocity $\bar{u}_{b}$ over the hill crest of height $H$ and the kinetic viscosity $\nu$.

At the upper and lower walls, no-slip condition is set for the velocity, and Neumann condition for the pressure.
For the eddy viscosity the low Reynolds number boundary condition is chosen (i.\,e., $\nu_{t}$ at the wall face is zero).

\begin{figure}[!ht]
    \centering
    \includegraphics[width=0.49\textwidth]{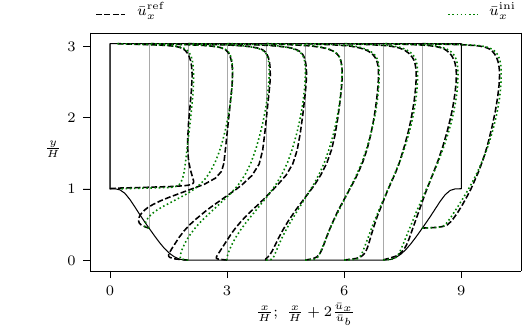}
    \caption{
        Stream wise velocity components of the reference $\bar{u}^{\mathrm{ref}}$, and the baseline RANS results $\bar{u}^{\mathrm{ini}}$ for the periodic hill setup at $\mathrm{Re} = \num{10595}$.
        All length scales are normalized by the hill height $H$ and all velocities by the bulk velocity at the hill crest $\bar{u}_{b}$, respectively.
    }
    \label{fig:ph_baseline_result}
\end{figure}

The resulting baseline RANS solution (cf. \cref{fig:ph_baseline_result}) is already close to the reference, except in the wake region of the hill.
This recirculation zone and the position of the reattachment point are not well captured, however.

\begin{figure}[!ht]
    \centering
    \includegraphics[width=0.49\textwidth]{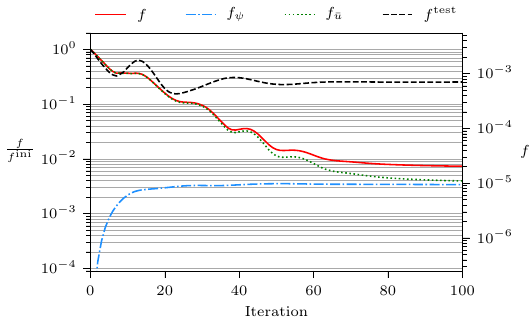}
    \hfill
    \includegraphics[width=0.49\textwidth]{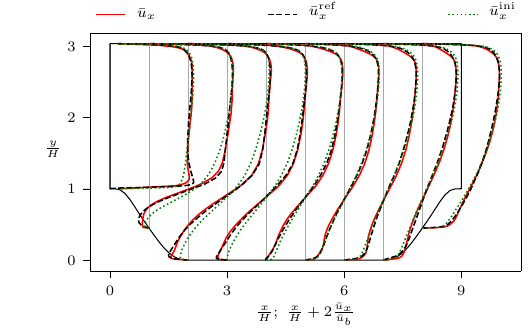}
    \caption{
        Data assimilation results with parameter $\psi_{k}$ for the periodic hill setup with the \emph{sparse} reference data distribution (cf. \cref{fig:ph_geometry_ref_map}).
        Shown are the cost function components and the test function (left), as well as the stream wise velocity components of the reference $\bar{u}^{\mathrm{ref}}$, the baseline RANS results $\bar{u}^{\mathrm{ini}}$, and the optimized results $\bar{u}$ (right).
    }
    \label{fig:ph_psi_results}
\end{figure}

The optimization is performed with a learning rate of $\eta=\num{8e-06}$, a number of total iteration steps of $N^{\mathrm{opt}}=\num{100}$, and a regularization weight of $w^{\mathrm{reg}}=\num{9e-04}$.
\Cref{fig:ph_psi_results} shows a clear reduction in both the cost and test function values, which can also be observed in the obtained velocity profiles.
Across the domain the optimized velocity field matches the reference very well with a particular improvement in the wake region.
Considering the low number of reference data points, this is a convincing improvement.
The reattachment point of the optimized result is at $x/H=\num{3.901}$ which is considerably closer to the reference value of $x/H=\num{4.268}$ than the baseline RANS result at $x/H=\num{3.069}$.

Results for denser reference data configurations are presented in \cref{sec:app_ph_psi}; similar levels of improvement can be observed.
Again, the same learning rate and regularization weights were employed for all configurations.

In \cref{sec:app_ph_alpha} equivalent results are presented for the tuning of parameter $\alpha$.
Although convincing results for this setup and approach have been presented in the literature using a more refined mesh (cf.~\cite{brenner22}), the results of this work suggest that this does not apply to the rather coarse meshes used here.
The corrective forcing term optimizations perform significantly better.

\subsubsection{Bump}

\begin{figure}[!ht]
    \centering
    \includegraphics{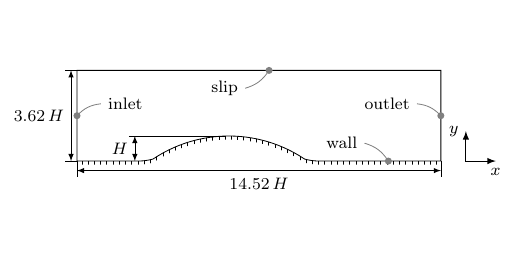}
    \hfill
    \includegraphics[trim=0 0 0 10,clip]{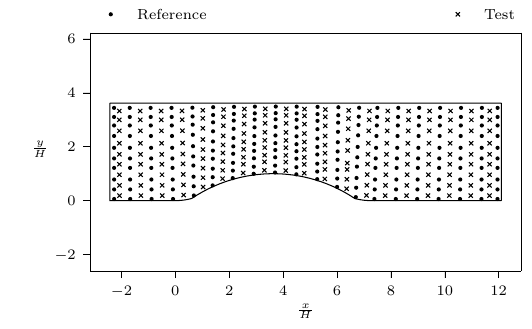}
    \caption{
        The simulation domain of the bump setup with indicated boundary types is shown in the left plot.
        Mean flow is in $x$ direction and all length scales are normalized by the bump height $H$.
        The distribution of \emph{reference} (dots) and \emph{test} (crosses) data points are shown in the right plot.
    }
    \label{fig:bump_geometry_ref_map}
\end{figure}

Another similar setup featuring a distinct but small separation bubble is the bump case by Webster~\etal.~\cite{webster96}, for which Matai and Durbin~\cite{matai19} provide averaged LES reference data.
In this analysis we focus on the specific bump configuration with height $H=\SI{42}{mm}$ as used in the original publication.
The Reynolds number of the flow is
\begin{equation}
    \mathrm{Re}
    =
    \frac{\bar{u}_{\infty} H}{\nu}
    =
    \num{28230}
\end{equation}
with viscosity $\nu$ and nominal free stream velocity $\bar{u}_{\infty}$.
The geometry with boundary conditions and the reference data distribution are shown in \cref{fig:bump_geometry_ref_map}.
For the upper boundary, a slip condition is chosen for all fields, while a no-slip condition is chosen for the velocity of the lower wall with wall functions for $k$, $\varepsilon$, and $\nu_{t}$ and a Neumann condition for pressure.
The velocity profile at the inlet is interpolated from the reference data, and the $k$, $\varepsilon$, and $\nu_{t}$ fields are assigned fixed values based on the free stream turbulence intensity of $\SI{0.2}{\percent}$ as reported in~\cite{webster96}.
For the outlet, a Neumann boundary condition is applied for all properties except pressure, for which a Dirichlet condition is used.
A coarse mesh of \num{15} cells in $x$-direction by \num{50} cells in $y$-direction (total cell count is \num{7500}) is used with approximately uniformly spaced, interlaced patterns of \num{200} reference and \num{171} test points.

\begin{figure}[!ht]
    \centering
    \includegraphics[width=0.49\textwidth]{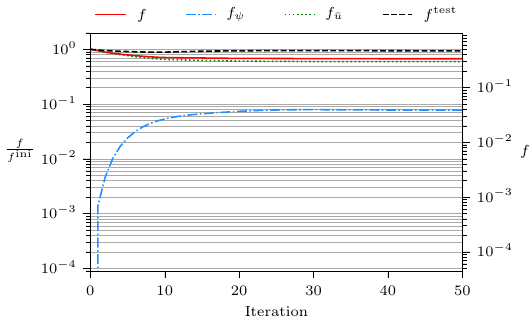}
    \hfill
    \includegraphics[width=0.49\textwidth]{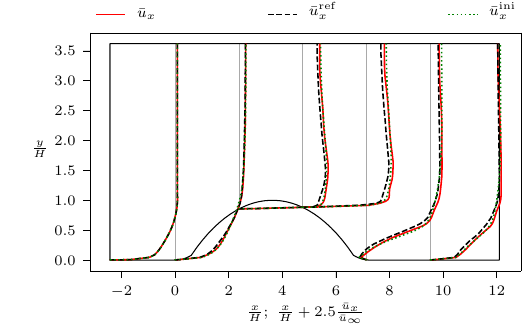}
    \caption{
        Data assimilation results with parameter $\psi_{k}$ for the bump setup (cf. \cref{fig:bump_geometry_ref_map}).
        Shown are the cost function components and the test function (left), as well as the stream wise velocity components of the reference $\bar{u}^{\mathrm{ref}}$, the baseline RANS results $\bar{u}^{\mathrm{ini}}$, and the optimized results $\bar{u}$ (right).
        All length scales are normalized by the bump height $H$ and all velocities by the free stream velocity $\bar{u}_{\infty}$, respectively.
    }
    \label{fig:bump_psi_results}
\end{figure}

Results for data assimilation in the bump case with parameter $\psi_{k}$ are presented in \cref{fig:bump_psi_results}.
From the velocity profile plots it is evident that the baseline RANS solution agrees very well with the reference data with only small deviations present around and downstream of the bump.
Data assimilation was performed with a learning rate of $\eta=\num{1e-03}$, a number of total iteration steps of $N^{\mathrm{opt}}=\num{50}$, and a regularization weight of $w^{\mathrm{reg}}=\num{2e-04}$.
The improvement of cost and test functions is clearly less than one order of magnitude.
Regarding the velocity profile plot, which only represent a small sample of the domain, no clear improvement can be seen.

Equivalent studies were performed for the tuning of $\alpha$ and $\nu_{t}$, respectively.
Neither of these two methods yielded an improvement in the cost function across a wide range of optimization and regularization parameter values.
However, while not achieving the same level of improvement observed in other cases, the corrective forcing approach stood out as the only method that showed any improvement in this apparently challenging setup.
Compared to the the periodic hill and backward facing step setups, the initial RANS solution is already very close to the reference and leaves only little room for improvement.

\section{Conclusions and outlook}
\label{sec:conclusion}

This paper presents an efficient data assimilation approach for RANS simulations based on coarse meshes and a corrective forcing term.
The latter is computed from a vector potential when only velocity reference data are used, as in this work, resulting in its divergence-free property.
While applicable in a general context, this approach shows particular advantages in two-dimensional setups where the parameter field simplifies to a scalar field.
The effect of the data assimilation parameter extends beyond the constraints of the eddy viscosity and is observed to consistently produce better optimized outcomes in comparison to alternative methods that rely on adjusting the eddy viscosity.
Even in the challenging case of flow over a bump, where the other methods failed entirely, the divergence-free forcing approach delivers a modest improvement.

The efficacy of the approach was established across three setups, namely flow over a backward facing step, periodic hills, and a bump, respectively.
In the backward facing step and periodic hill setups, different reference data distributions were considered.
It has been observed that the regularization weight parameters 
As anticipated, more tightly packed reference distributions yielded superior outcomes, but regularization allowed for accurate results with sparse distributions.
It has been observed that the same values for the regularization weight could be used for different reference data distributions of a specific setup.
While a number of flow setups and reference data distributions were considered in this work, the optimal number and placement of reference and test data points for this data assimilation approach still needs to be further investigated.
Ideally, only very little few well placed reference data points need to be sourced in an experimental context.
This could be combined with a more in-depth analysis of parameter field regularization and fine-tuning of the regularization weight parameter.

The required number of optimization steps in the presented cases is reasonable, as the computational cost of the incremental flow solutions is relatively low.
A more detailed study of alternative optimization methods could be performed to further improve the overall performance of the approach.


\section*{Acknowledgments}

The authors would like to thank Roberto Iafigliola for his contributions during student projects and as student research assistant and Robert Epp for fruitful discussions.
Financial support from the ETH Foundation is gratefully acknowledged.


\clearpage

\appendix

\gdef\thesection{\Alph{section}}
\makeatletter
\renewcommand\@seccntformat[1]{\appendixname\ \csname the#1\endcsname.\hspace{0.5em}}
\makeatother

\section{Discrete adjoint method}
\label{sec:app_adjoint}

To derive an expression for the cost function gradient, we introduce a Lagrangian $\mathcal{L}$ as in \cref{eq:lagrangian} with Lagrangian multiplier $\lambda$ (cf.\cref{eq:lambda}).
The corresponding gradient with respect to the parameters $\xi$ is derived as
\begin{align}
    \label{eq:app_adjoint_gradient}
    \frac{\mathrm{d} f}{\mathrm{d} \xi}
    &=
    \nonumber
    \frac{\mathrm{d} \mathcal{L}}{\mathrm{d} \xi}\\
    &=
    \nonumber
    \frac{\mathrm{d} f}{\mathrm{d} \xi}
    -
    \lambda^{T} \frac{\mathrm{d} R}{\mathrm{d} \xi}
    -
    \frac{\mathrm{d} \lambda^{T}}{\mathrm{d} \xi} R\\
    \nonumber
    &=
    \frac{\partial f_{\xi}}{\partial \xi}
    +
    \frac{\partial f_{U}}{\partial U} \frac{\partial U}{\partial \xi}
    -
    \lambda^{T} \frac{\partial R}{\partial \xi}
    -
    \lambda^{T} \frac{\partial R}{\partial U} \frac{\partial U}{\partial \xi}\\
    \nonumber
    &=
    \frac{\partial f_{\xi}}{\partial \xi}
    -
    \lambda^{T} \frac{\partial R}{\partial \xi}
    +
    \Biggl[
        \underbrace{
            \frac{\partial f_{U}}{\partial U}
            - \lambda^{T} \frac{\partial R}{\partial U}
        }_{=0}
    \Biggr]
    \frac{\partial U}{\partial \xi}\\
    &=
    \frac{\partial f_{\xi}}{\partial \xi}
    -
    \lambda^{T} \frac{\partial R}{\partial \xi}
    \, .
\end{align}
To simplify the second to last line above, the Lagrangian multiplier $\lambda$ is chosen such that the expression in brackets vanishes.
Rearranging the terms and considering that the derivative of the linear forward problem residual $R$ with respect to the linear forward problem solution $U$ corresponds to the respective system matrix $\mathbf{A}_{U}$ yields
\begin{equation}
    \label{eq:app_coupled_adjoint}
    \mathbf{A}_{U}^{T} \, \lambda
    =
    \frac{\partial f_{U}}{\partial U}^{T}
    \, .
\end{equation}
The right-hand side of this equation is analytically derived from the cost function and thus comes at low computational cost.
Solving this system of linear equations is comparable in cost to solving the forward system for $U$ and independent of the number of parameters $\xi$.

The derivatives of the forward problem residual with respect to the parameters $\frac{\partial R}{\partial \xi}$ are evaluated using the approximate and efficient approach introduced by Brenner~\etal.~\cite{brenner22}.
In particular, the forward problem $R$ is linearized with respect to $\xi$, denoted as $R_{\xi}$, in \emph{OpenFOAM} as
\begin{equation}
    R_{\xi}
    =
    \mathbf{A}_{\xi} \, \xi
    -
    b_{\xi}
    =
    0
    \, .
\end{equation}
The derivatives thus correspond to the system matrix $\mathbf{A}_{\xi}$.
Combining these approximate terms with \cref{eq:app_adjoint_gradient} and the solution of \cref{eq:app_coupled_adjoint} and expanding the terms yields
\begin{equation}
    \frac{\mathrm{d} f}{\mathrm{d} \xi}
    =
    \frac{\partial f_{\xi}}{\partial \xi}
    -
    \lambda_{\bar{u}_{x}}^{T} \mathbf{A}_{\bar{u}_{x},\xi}
    -
    \lambda_{\bar{u}_{y}}^{T} \mathbf{A}_{\bar{u}_{y},\xi}
    -
    \lambda_{\bar{u}_{z}}^{T} \mathbf{A}_{\bar{u}_{z},\xi}
    -
    \lambda_{\bar{p}}^{T} \mathbf{A}_{\bar{p},\xi}
\end{equation}
for the discrete adjoint gradient in RANS.

\section{Improved modified eddy-viscosity approach}
\label{sec:app_alpha}

The RANS data assimilation approach by Brenner~\etal.~\cite{brenner22} is based on the momentum equation
\begin{equation}
    \label{eq:momentum_alpha}
    \frac{\partial \bar{u}_{i} \bar{u}_{j}}{\partial x_{j}}
    +
    \frac{\partial p^{*}}{\partial x_{i}}
    -
    \frac{\partial}{\partial x_{j}}
    \left[
        2
        \left(
            \nu
            +
            \alpha
            \nu_{t}^{b}
        \right)
        \bar{S}_{ij}
    \right]
    =
    0
    \, ,
\end{equation}
with modified pressure
\begin{equation}
    p^{*}
    =
    \frac{\bar{p}}{\rho}
    +
    \frac{2}{3} k
    \, ,
\end{equation}
and baseline eddy viscosity $\nu_{t}^{b}$, which is kept constant (i.\,e. \emph{frozen}) for the data assimilation procedure.
Instead, the spatially averaged parameter field $\alpha$ is tuned in the process, starting from an initially uniform field of $\alpha^{\left(0\right)}=\num{1}$.
Brenner~\etal.~\cite{brenner22} showed good performance of this method for the periodic hill setup, despite its limitations to only influence the flow in regions with sufficiently large eddy viscosity and velocity gradient values.

The linearized version of \cref{eq:momentum_alpha} reads
\begin{equation}
    \bar{\phi}_{j}\frac{\partial \bar{u}_{i} }{\partial x_{j}}
    -
    \frac{\partial}{\partial x_{j}}
    \left[
        \left(
            \nu
            +
            \alpha
            \nu_{t}^{b}
        \right)
        \frac{\partial \bar{u}_{i}}{\partial x_{j}}
    \right]
    =
    -\frac{\partial p^{\dag}}{\partial x_{i}}
    +
    \frac{\partial}{\partial x_{j}}
    \left[
        \left(
            \nu
            +
            \alpha
            \nu_{t}^{b}
        \right)
        \frac{\partial \bar{u}_{j}}{\partial x_{i}}
    \right]
    \, ,
\end{equation}
with linearized velocity $\bar{\phi}_{j}$ based on the last iteration step; terms considered \emph{implicitly} in the discretized system of equations are grouped on the left-hand side of the equation, while \emph{explicitly} treated terms are grouped on the right-hand side.

In the present work, the solver was modified to provide periodic solutions $\lambda$ of the adjoint system of equations.
This mandated additional steps to improve solver convergence and stability.
The parameter updates based on the periodic adjoint gradients, and thus the parameter fields, are consequently periodic as well.
Accordingly, the TV regularization implementation was extended to consider cell boundaries across periodic domains.

A minor change was introduced to the cost and test functions, which are now normalized by the volume and not the number of reference and test cells, respectively.
In addition, the TV regularization is now considered without the extra smoothing step that was part of the original approach and consisted of the solution of an additional Helmholtz equation.
This is sufficient to provide smooth results and further reduces the computational cost.

\section{Implicit curl operator: Implementation and validation}
\label{sec:app_imp_curl}

The \emph{foam-extend-5.0}~\cite{fe50} package does not provide an \emph{implicit} curl operator that is required to evaluate the adjoint gradient for DA parameter $\psi_{k}$ (cf. \cref{eq:dRdPsi}).
As part of this work, this relevant part of the missing operator was successfully implemented, validated, and employed in the DA procedure.
The implementation of the operator is presented next, followed by a description of the validation approach, and a summary of the validation result.

\subsection{Implementation}

The implementation does not feature a full-fledged implicit \emph{OpenFOAM} operator for the curl of a vector field, but only the part relevant to this work, i.\,e., the corresponding system matrix.
Applying the curl operator on a vector field $v$ yields
\begin{equation}
    \label{eq:app_curl_impl}
    \nabla
    \times
    v
    =
    \begin{bmatrix}
        \frac{\partial}{\partial x}\\
        \frac{\partial}{\partial y}\\
        \frac{\partial}{\partial z}
    \end{bmatrix}
    \times
    \begin{bmatrix}
        v_{x}\\
        v_{y}\\
        v_{z}
    \end{bmatrix}
    =
    \begin{bmatrix}
        0 & -\frac{\partial}{\partial z} & \frac{\partial}{\partial y}\\
        \frac{\partial}{\partial z} & 0 & -\frac{\partial}{\partial x}\\
        -\frac{\partial}{\partial y} & \frac{\partial}{\partial x} & 0
    \end{bmatrix}
    \begin{bmatrix}
        v_{x}\\
        v_{y}\\
        v_{z}
    \end{bmatrix}
    \approx
    \mathbf{A}v
    -
    b
    \, ,
\end{equation}
where the last step represents the discrete representation.
This representation bears some similarities with the gradient operator applied to a scalar field $s$ as
\begin{equation}
    \label{eq:app_grad_impl}
    \nabla
    s
    =
    \begin{bmatrix}
        \frac{\partial}{\partial x}\\
        \frac{\partial}{\partial y}\\
        \frac{\partial}{\partial z}
    \end{bmatrix}
    s
    \, .
\end{equation}
Specifically, the operator matrix entries in \cref{eq:app_curl_impl} correspond to the entries in the operator in \cref{eq:app_grad_impl}.
In the implementation this is leveraged by assembling the implicit curl operator system matrix with the corresponding elements of the implicit gradient operator according to \cref{eq:app_curl_impl}.

\subsection{Validation}

The \emph{method of manufactured solution} is employed to validate the implementation.
To this end, an analytical reference solution $v^{\mathrm{ref}}$ is defined and evaluated based on explicit operators to obtain a value for the right-hand side of the system of equations.
The operator implementation is then validated by solving this system of equations for $v$ and comparing the result to the reference field $v^{\mathrm{ref}}$

The implicit curl operator produces only off-diagonal matrix entries, so a system of equations consisting of only this implicit operator alone cannot be solved numerically.
To obtain a better suited matrix, an additional term $a v$ with parameter $a$ is introduced to give more weight to the matrix diagonal by
\begin{equation}
    \underbrace{
        a v
        +
        \nabla \times v
    }_{\mathrm{implicit}}
    =
    \underbrace{
        a v^{\mathrm{ref}}
        +
        \nabla \times v^{\mathrm{ref}}
    }_{\mathrm{explicit}}
    \, ,
\end{equation}
where the \emph{implicit} left-hand side corresponds to matrix $\boldsymbol{A}$ of the system of linear equations and the \emph{explicit} right-hand side to the source term $r$.
In discrete form, this reads
\begin{equation}
    \label{eq:app_curl_discrete}
    \boldsymbol{A} v
    =
    r
    \, .
\end{equation}

For validation, an analytical reference solution $v^{\mathrm{ref}}$ was selected and evaluated at the cell center locations of the test case meshes, using the implicit source and curl operators provided by \emph{OpenFOAM}.
Accordingly, the analytical expression for the right-hand side was derived and evaluated.
A total of seven three-dimensional test cases were set up, which differ in boundary conditions, domain shape and size, mesh resolution, and mesh cell properties.
In particular, periodic and Dirichlet boundary conditions were considered, and the mesh cell shapes ranged from uniformly hexahedral to very elongated and skewed.
\Cref{eq:app_curl_discrete} was then solved numerically for all cases, where relatively large values had to be chosen for $a$ to reach convergence.

From the results of all test cases it was found, that there are no boundary effects and that the results agree very well with the reference.
The implementation of the implicit curl operator is thus concluded to be correct.

\section{Additional periodic hill results for the divergence-free source term approach}
\label{sec:app_ph_psi}

\begin{figure}[!ht]
    \centering
    \includegraphics[width=0.49\textwidth,trim=0 0 0 10,clip]{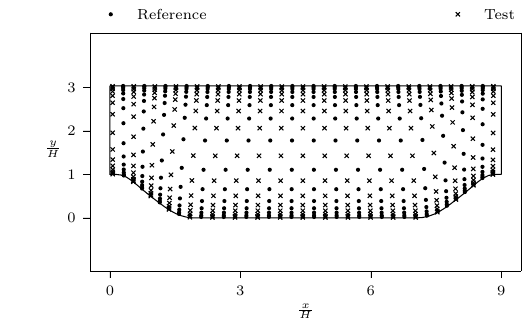}
    \hfill
    \includegraphics[width=0.49\textwidth,trim=0 0 0 10,clip]{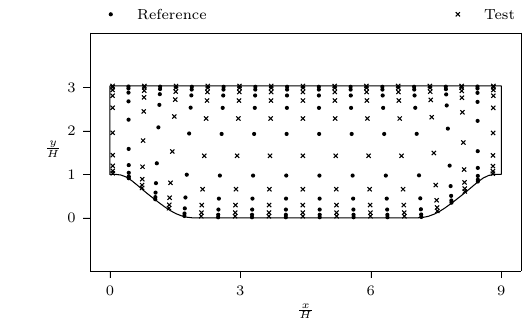}
    \caption{
        Distribution of \emph{reference} (dots) and \emph{test} (crosses) data points for the \emph{dense} (left) and \emph{intermediate} (right) configurations of the periodic hill setup (cf. \cref{fig:ph_geometry_ref_map}).
    }
    \label{fig:app_ph_ref_map}
\end{figure}

In this section, additional results for the periodic hill case from \cref{subsubsec:res_ph} are presented.
In particular, two additional sets of reference and test data distributions are introduced, as depicted in \cref{fig:app_ph_ref_map}.
These setups are referred to as \emph{dense} and \emph{intermediate} reference density cases, respectively.
They feature the same mesh and boundary conditions as the \emph{sparse} case in \cref{subsubsec:res_ph}, which has \num{48} reference and \num{42} test points.
In comparison, the \emph{intermediate} case features \num{120} reference and \num{130} test points, while the corresponding numbers of the \emph{dense} case are \num{270} reference and \num{285} test points, respectively.

To facilitate the analysis of the influence of the number of reference data points on the optimization result, the optimization and regularization parameters are chosen identically for all three cases.
The learning rate $\eta$ was set to \num{8e-06}, a number of total iteration steps of $N^{\mathrm{opt}}=\num{100}$, and the regularization weight $w^{\mathrm{reg}}$ to \num{9e-04}.

\begin{figure}[!ht]
    \centering
    \includegraphics[width=0.49\textwidth,trim=0 22 0 0,clip]{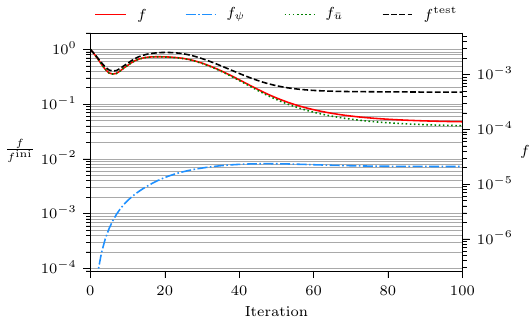}
    \hfill
    \includegraphics[width=0.49\textwidth,trim=0 22 0 0,clip]{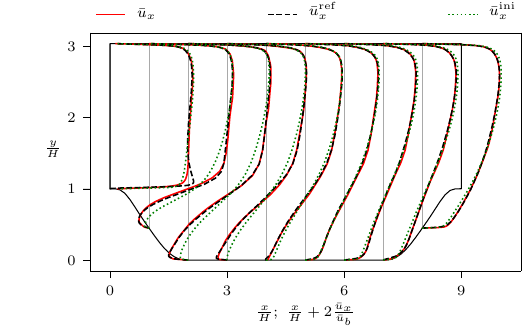}\\
    \includegraphics[width=0.49\textwidth,trim=0 22 0 13,clip]{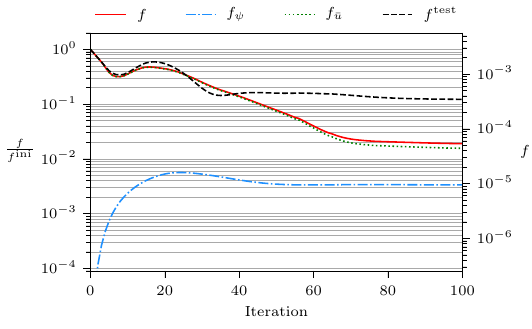}
    \hfill
    \includegraphics[width=0.49\textwidth,trim=0 22 0 13,clip]{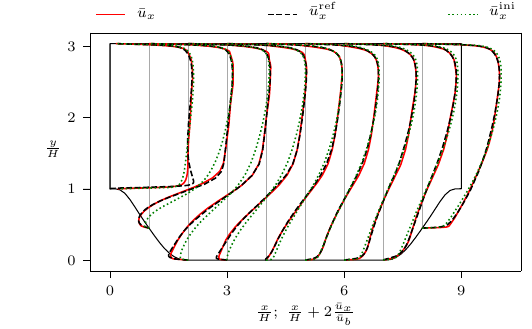}\\
    \includegraphics[width=0.49\textwidth,trim=0 0 0 13,clip]{PHZPsiSparse_cost_function.pdf}
    \hfill
    \includegraphics[width=0.49\textwidth,trim=0 0 0 13,clip]{PHZPsiSparse_profile_U_URef_UIni.pdf}
    \caption{
        Cost function (left) and mean stream wise velocity profiles of the mean-flow direction component $\bar{u}_{x}$ (right) for data assimilation parameter $\psi_{k}$ in the periodic hill geometry.
        The rows correspond to the \emph{dense} (top), \emph{intermediate} (center), and \emph{sparse} (bottom) reference patterns, respectively (cf. \cref{fig:app_ph_ref_map,fig:ph_geometry_ref_map}).
        Plots for the \emph{sparse} configuration are identical to those presented in \cref{fig:ph_psi_results}.
    }
    \label{fig:app_ph_psi_results}
\end{figure}

\Cref{fig:app_ph_psi_results} depicts the resulting velocity profiles and the cost function plots for the \emph{dense}, \emph{intermediate}, and \emph{sparse} reference data.
Qualitatively comparing the optimized velocity profiles shows clear improvements in the recirculation region, where the difference between baseline RANS $\bar{u}^{\mathrm{ini}}$ and reference $\bar{u}^{\mathrm{ref}}$ is the greatest in the hill wake for all setups.
The main deviations between the optimized results and the reference are observed near the lower wall in the inlet plane, with other smaller deviations just below the centerline.
These are attributed to the reference data distribution, as there is a gap between the first plane of reference locations near the inlet and outlet planes, respectively.
Similarly, the density of reference data points is closer in the near wall region than in the bulk region.

The reattachment length of the optimized results are closer to the reference values, as reported in \cref{tab:ph_psi_reattachment}.
This is consistent with observations made in the velocity profile plots.

\begin{table}[!ht]
    \centering
    \caption{
        This table compares the reattachment lengths of the baseline RANS solution and the data assimilation results for parameter $\psi_{k}$ with different reference data configurations.
        The reattachment lengths are reported as the distance from the left boundary normalized by the hill crest height $H$.
        The relative errors are reported with respect to the reference value.
        }
    \label{tab:ph_psi_reattachment}
    \begin{tabular}{lrrrrr}
        \toprule
                                       & Reference   & Baseline             & Dense               & Intermediate        & Sparse\\
        \midrule
        Normalized reattachment length & \num{4.268} & \num{3.069}          & \num{3.854}         & \num{3.933}         & \num{3.915}\\
        Relative error                 &           - & \SI{28.10}{\percent} & \SI{9.71}{\percent} & \SI{7.86}{\percent} & \SI{8.28}{\percent}\\
        \bottomrule
    \end{tabular}
\end{table}

\section{Periodic hill results for the modified eddy viscosity approach}
\label{sec:app_ph_alpha}

For the same three periodic hill setups described in \cref{sec:app_ph_psi}, data assimilation was performed for parameter $\alpha$.
These cases are comparable to the ones presented in \cite{brenner22}, albeit with all the modifications discussed in \cref{sec:app_alpha}.
Specifically, the periodicity of the Lagrangian multipliers $\lambda$ is considered, which requires an under-relaxation of the corresponding system of linear equations for a stable solution process.
This is also true for the periodic hill results with the divergence-free forcing term presented in \cref{sec:app_ph_psi}.
Further, the mesh in this work has a much lower resolution of only \num{4400} cells, instead of \num{23400} in~\cite{brenner22}; the Demon-Adam optimizer is used over the Gauss-Newton method, and no additional smoothing step is employed.

\begin{figure}[!ht]
    \centering
    \includegraphics[width=0.49\textwidth,trim=0 22 0 0,clip]{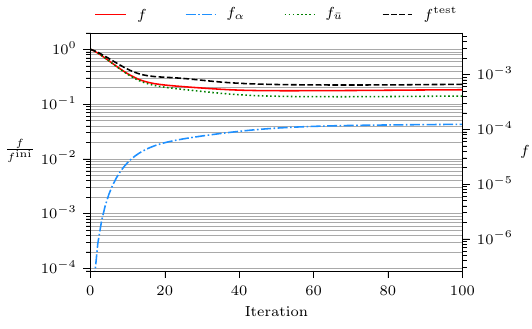}
    \hfill
    \includegraphics[width=0.49\textwidth,trim=0 22 0 0,clip]{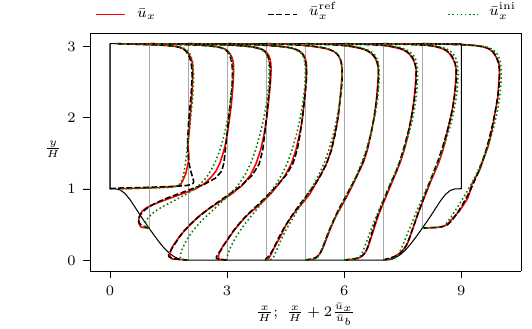}\\
    \includegraphics[width=0.49\textwidth,trim=0 22 0 13,clip]{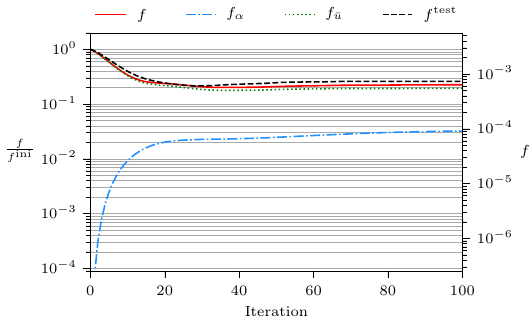}
    \hfill
    \includegraphics[width=0.49\textwidth,trim=0 22 0 13,clip]{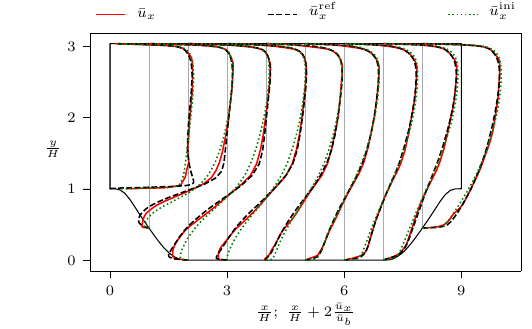}\\
    \includegraphics[width=0.49\textwidth,trim=0 0 0 13,clip]{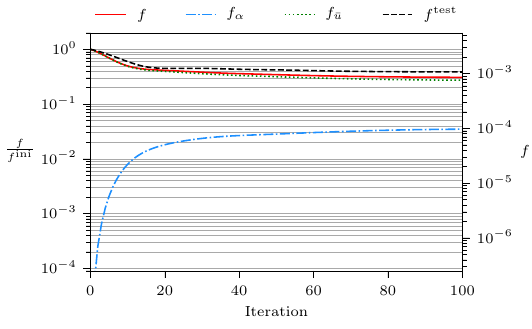}
    \hfill
    \includegraphics[width=0.49\textwidth,trim=0 0 0 13,clip]{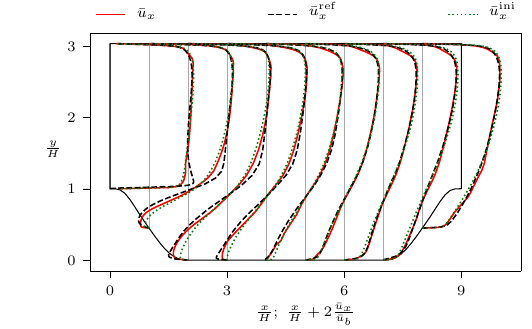}
    \caption{
        Cost function (left) and mean stream wise velocity profiles of the mean-flow direction component $\bar{u}_{x}$ (right) for data assimilation parameter $\alpha$ in the periodic hill geometry.
        The rows correspond to the \emph{dense} (top), \emph{intermediate} (center), and \emph{sparse} (bottom) reference patterns (cf. \cref{fig:ph_geometry_ref_map,fig:app_ph_ref_map}).
    }
    \label{fig:app_ph_alpha_results}
\end{figure}

Data assimilation was performed with a learning rate of $\eta=\num{1e-03}$, a number of total iteration steps of $N^{\mathrm{opt}}=\num{100}$, and a regularization weight of $w^{\mathrm{reg}}=\num{5e-08}$.
\Cref{fig:app_ph_alpha_results} shows the resulting cost function and velocity profile plots.
An improvement in cost function values and velocity profiles is observed for all three configurations, while the results are better for higher numbers of reference data points.
However, both the reduction in cost and test functions and the improvement in velocity profiles are less than in the results for the divergence-free forcing approach.

Compared with the literature results in \cite{brenner22}, the performance here is worse.
This is attributed to the vastly reduced mesh resolution and the lack of an additional smoothing step which requires the usage of stronger regularization.
For the \emph{sparse} configuration, the regularization almost completely blocks any improvement.

As reported for the $\psi_{k}$ parameter, the reattachment lengths of the optimized results are also closer to the reference values for the $\alpha$ optimization parameter (cf. \cref{tab:ph_alpha_reattachment}).

\begin{table}[!ht]
    \centering
    \caption{
        This table compares the reattachment lengths of the baseline RANS solution and the data assimilation results for parameter $\alpha$ with different reference data configurations.
        The reattachment lengths are reported as the distance from the left boundary normalized by the hill crest height $H$.
        The relative errors are reported with respect to the reference value.
        }
    \label{tab:ph_alpha_reattachment}
    \begin{tabular}{lrrrrr}
        \toprule
                                       & Reference   & Baseline             & Dense                & Intermediate         & Sparse\\
        \midrule
        Normalized reattachment length & \num{4.268} & \num{3.070}          & \num{3.875}          & \num{4.486}          & \num{3.984}\\
        Relative error                 &           - & \SI{28.07}{\percent} & \SI{9.22}{\percent}  & \SI{5.08}{\percent}  & \SI{6.67}{\percent}\\
        \bottomrule
    \end{tabular}
\end{table}



\clearpage
\begin{thebibliography}{10}
\expandafter\ifx\csname url\endcsname\relax
  \def\url#1{\texttt{#1}}\fi
\expandafter\ifx\csname urlprefix\endcsname\relax\def\urlprefix{URL }\fi
\expandafter\ifx\csname href\endcsname\relax
  \def\href#1#2{#2} \def\path#1{#1}\fi

\bibitem{spalart16}
P.~R. Spalart, V.~Venkatakrishnan, On the role and challenges of {CFD} in the aerospace industry, The Aeronautical Journal 120~(1223) (2016) 209--232.
\newblock \href {https://doi.org/10.1017/aer.2015.10} {\path{doi:10.1017/aer.2015.10}}.

\bibitem{stopford02}
P.~J. Stopford, Recent applications of {CFD} modelling in the power generation and combustion industries, Applied Mathematical Modelling 26~(2) (2002) 351--374.
\newblock \href {https://doi.org/10.1016/s0307-904x(01)00066-x} {\path{doi:10.1016/s0307-904x(01)00066-x}}.

\bibitem{costes12}
M.~Costes, T.~Renaud, B.~Rodriguez, Rotorcraft simulations: a challenge for {CFD}, International Journal of Computational Fluid Dynamics 26~(6-8) (2012) 383--405.
\newblock \href {https://doi.org/10.1080/10618562.2012.726710} {\path{doi:10.1080/10618562.2012.726710}}.

\bibitem{lain22}
S.~Lain, O.~D.~L. Mejia, Special issue on ``{CFD} modelling and simulation of water turbines'', Processes 10~(11) (2022) 2410.
\newblock \href {https://doi.org/10.3390/pr10112410} {\path{doi:10.3390/pr10112410}}.

\bibitem{mani23}
M.~Mani, A.~J. Dorgan, A perspective on the state of aerospace computational fluid dynamics technology, Annual Review of Fluid Mechanics 55~(1) (2023) 431--457.
\newblock \href {https://doi.org/10.1146/annurev-fluid-120720-124800} {\path{doi:10.1146/annurev-fluid-120720-124800}}.

\bibitem{wilcox06}
D.~C. Wilcox, Turbulence modeling for {CFD}, no. Bd. 1 in Turbulence Modeling for CFD, DCW Industries, 2006.

\bibitem{pope00}
S.~B. Pope, Turbulent flows, seventh printing edition Edition, Cambridge University Press, 2010.
\newblock \href {https://doi.org/10.1017/cbo9780511840531} {\path{doi:10.1017/cbo9780511840531}}.

\bibitem{menter94}
F.~R. Menter, Two-equation eddy-viscosity turbulence models for engineering applications, AIAA Journal 32~(8) (1994) 1598--1605.
\newblock \href {https://doi.org/10.2514/3.12149} {\path{doi:10.2514/3.12149}}.

\bibitem{spalart92}
P.~Spalart, A.~Allmaras, A one-equation turbulence model for aerodynamic flows, in: 30th Aerospace Sciences Meeting and Exhibit, American Institute of Aeronautics and Astronautics, 1992, p. 439.
\newblock \href {https://doi.org/10.2514/6.1992-439} {\path{doi:10.2514/6.1992-439}}.

\bibitem{craft96}
T.~J. Craft, B.~E. Launder, K.~Suga, Development and application of a cubic eddy-viscosity model of turbulence, International Journal of Heat and Fluid Flow 17~(2) (1996) 108--115.
\newblock \href {https://doi.org/10.1016/0142-727x(95)00079-6} {\path{doi:10.1016/0142-727x(95)00079-6}}.

\bibitem{duraisamy18}
K.~Duraisamy, G.~Iaccarino, H.~Xiao, Turbulence modeling in the age of data, Annual Review of Fluid Mechanics 51~(1) (2018) 357--377.
\newblock \href {https://doi.org/10.1146/annurev-fluid-010518-040547} {\path{doi:10.1146/annurev-fluid-010518-040547}}.

\bibitem{xiao17}
H.~Xiao, J.-X. Wang, R.~G. Ghanem, A random matrix approach for quantifying model-form uncertainties in turbulence modeling, Computer Methods in Applied Mechanics and Engineering 313 (2017) 941--965.
\newblock \href {https://doi.org/10.1016/j.cma.2016.10.025} {\path{doi:10.1016/j.cma.2016.10.025}}.

\bibitem{xiao19a}
H.~Xiao, P.~Cinnella, Quantification of model uncertainty in {RANS} simulations: a review, Progress in Aerospace Sciences 108 (2019) 1--31.
\newblock \href {https://doi.org/10.1016/j.paerosci.2018.10.001} {\path{doi:10.1016/j.paerosci.2018.10.001}}.

\bibitem{karniadakis21}
G.~E. Karniadakis, I.~G. Kevrekidis, L.~Lu, P.~Perdikaris, S.~Wang, L.~Yang, Physics-informed machine learning, Nature Reviews Physics (May 2021).
\newblock \href {https://doi.org/10.1038/s42254-021-00314-5} {\path{doi:10.1038/s42254-021-00314-5}}.

\bibitem{wang17a}
J.-X. Wang, J.-L. Wu, H.~Xiao, Physics-informed machine learning approach for reconstructing {R}eynolds stress modeling discrepancies based on {DNS} data, Physical Review Fluids 2~(3) (Mar. 2017).
\newblock \href {https://doi.org/10.1103/physrevfluids.2.034603} {\path{doi:10.1103/physrevfluids.2.034603}}.

\bibitem{raissi18}
M.~Raissi, P.~Perdikaris, G.~E. Karniadakis, Physics-informed neural networks: a deep learning framework for solving forward and inverse problems involving nonlinear partial differential equations, Journal of Computational Physics (Nov. 2018).
\newblock \href {https://doi.org/10.1016/j.jcp.2018.10.045} {\path{doi:10.1016/j.jcp.2018.10.045}}.

\bibitem{sliwinski23}
L.~Sliwinski, G.~Rigas, Mean flow reconstruction of unsteady flows using physics-informed neural networks, Data-Centric Engineering 4 (2023).
\newblock \href {https://doi.org/10.1017/dce.2022.37} {\path{doi:10.1017/dce.2022.37}}.

\bibitem{kwon18}
I.-H. Kwon, S.~English, W.~Bell, R.~Potthast, A.~Collard, B.~Ruston, Assessment of progress and status of data assimilation in numerical weather prediction, Bulletin of the American Meteorological Society 99~(5) (2018) ES75--ES79.
\newblock \href {https://doi.org/10.1175/bams-d-17-0266.1} {\path{doi:10.1175/bams-d-17-0266.1}}.

\bibitem{koltukluoglu18}
T.~S. Koltukluoğlu, P.~J. Blanco, Boundary control in computational haemodynamics, Journal of Fluid Mechanics 847 (2018) 329--364.
\newblock \href {https://doi.org/10.1017/jfm.2018.329} {\path{doi:10.1017/jfm.2018.329}}.

\bibitem{epp20}
R.~Epp, F.~Schmid, B.~Weber, P.~Jenny, Predicting vessel diameter changes to up-regulate biphasic blood flow during activation in realistic microvascular networks, Frontiers in Physiology 11 (Oct. 2020).
\newblock \href {https://doi.org/10.3389/fphys.2020.566303} {\path{doi:10.3389/fphys.2020.566303}}.

\bibitem{averweg22}
S.~Averweg, A.~Schwarz, C.~Schwarz, J.~Schröder, {3D} modeling of generalized {Newtonian} fluid flow with data assimilation using the least-squares finite element method, Computer Methods in Applied Mechanics and Engineering 392 (2022) 114668.
\newblock \href {https://doi.org/10.1016/j.cma.2022.114668} {\path{doi:10.1016/j.cma.2022.114668}}.

\bibitem{asch16}
M.~Asch, M.~Bocquet, M.~Nodet, Data assimilation: methods, algorithms, and applications, Society for Industrial and Applied Mathematics, Philadelphia, 2016.
\newblock \href {https://doi.org/10.1137/1.9781611974546} {\path{doi:10.1137/1.9781611974546}}.

\bibitem{bradley13}
A.~M. Bradley, \href{https://cs.stanford.edu/~ambrad/adjoint_tutorial.pdf}{{PDE}-constrained optimization and the adjoint method}, Report, Stanford University (Jun. 2013).
\newline\urlprefix\url{https://cs.stanford.edu/~ambrad/adjoint_tutorial.pdf}

\bibitem{yang15}
Y.~Yang, C.~Robinson, D.~Heitz, E.~Mémin, Enhanced ensemble-based {4DVar} scheme for data assimilation, Computers \& Fluids 115 (2015) 201--210.
\newblock \href {https://doi.org/10.1016/j.compfluid.2015.03.025} {\path{doi:10.1016/j.compfluid.2015.03.025}}.

\bibitem{hafez22}
A.~M. Hafez, A.~A. El-Rahman, H.~A. Khater, Field inversion for transitional flows using continuous adjoint methods, Physics of Fluids (Nov. 2022).
\newblock \href {https://doi.org/10.1063/5.0128522} {\path{doi:10.1063/5.0128522}}.

\bibitem{fleischli21}
B.~Fleischli, L.~Mangani, A.~D. Rio, E.~Casartelli, A discrete adjoint method for pressure-based algorithms, Computers \& Fluids (2021) 105037\href {https://doi.org/10.1016/j.compfluid.2021.105037} {\path{doi:10.1016/j.compfluid.2021.105037}}.

\bibitem{dilgen18}
C.~B. Dilgen, S.~B. Dilgen, D.~R. Fuhrman, O.~Sigmund, B.~S. Lazarov, Topology optimization of turbulent flows, Computer Methods in Applied Mechanics and Engineering 331 (2018) 363--393.
\newblock \href {https://doi.org/10.1016/j.cma.2017.11.029} {\path{doi:10.1016/j.cma.2017.11.029}}.

\bibitem{cato23}
A.~S. Cato, P.~S. Volpiani, V.~Mons, O.~Marquet, D.~Sipp, Comparison of different data-assimilation approaches to augment {RANS} turbulence models, Computers \& Fluids 266 (2023) 106054.
\newblock \href {https://doi.org/10.1016/j.compfluid.2023.106054} {\path{doi:10.1016/j.compfluid.2023.106054}}.

\bibitem{foures14}
D.~P.~G. Foures, N.~Dovetta, D.~Sipp, P.~J. Schmid, A data-assimilation method for {R}eynolds-averaged {N}avier--{S}tokes-driven mean~flow reconstruction, Journal of Fluid Mechanics 759 (2014) 404--431.
\newblock \href {https://doi.org/10.1017/jfm.2014.566} {\path{doi:10.1017/jfm.2014.566}}.

\bibitem{li22}
S.~Li, C.~He, Y.~Liu, A data assimilation model for wall pressure-driven mean flow reconstruction, Physics of Fluids 34~(1) (2022) 015101.
\newblock \href {https://doi.org/10.1063/5.0076754} {\path{doi:10.1063/5.0076754}}.

\bibitem{patel24}
Y.~Patel, V.~Mons, O.~Marquet, G.~Rigas, Turbulence model augmented physics-informed neural networks for mean-flow reconstruction, Physical Review Fluids 9~(3) (2024) 034605.
\newblock \href {https://doi.org/10.1103/physrevfluids.9.034605} {\path{doi:10.1103/physrevfluids.9.034605}}.

\bibitem{franceschini20a}
L.~Franceschini, D.~Sipp, O.~Marquet, Mean-flow data assimilation based on minimal correction of turbulence models: application to turbulent high {R}eynolds number backward-facing step, Physical Review Fluids 5~(9) (Sep. 2020).
\newblock \href {https://doi.org/10.1103/physrevfluids.5.094603} {\path{doi:10.1103/physrevfluids.5.094603}}.

\bibitem{brenner22}
O.~Brenner, P.~Piroozmand, P.~Jenny, Efficient assimilation of sparse data into {RANS}-based turbulent flow simulations using a discrete adjoint method, Journal of Computational Physics 471 (2022) 111667.
\newblock \href {https://doi.org/10.1016/j.jcp.2022.111667} {\path{doi:10.1016/j.jcp.2022.111667}}.

\bibitem{papadimitriou15}
D.~I. Papadimitriou, C.~Papadimitriou, Optimal sensor placement for the estimation of turbulence model parameters in {CFD}, International Journal for Uncertainty Quantification 5~(6) (2015) 545--568.
\newblock \href {https://doi.org/10.1615/int.j.uncertaintyquantification.2015015239} {\path{doi:10.1615/int.j.uncertaintyquantification.2015015239}}.

\bibitem{mons17}
V.~Mons, J.-C. Chassaing, P.~Sagaut, Optimal sensor placement for variational data assimilation of unsteady flows past a rotationally oscillating cylinder, Journal of Fluid Mechanics 823 (2017) 230--277.
\newblock \href {https://doi.org/10.1017/jfm.2017.313} {\path{doi:10.1017/jfm.2017.313}}.

\bibitem{epp22}
R.~Epp, F.~Schmid, P.~Jenny, Hierarchical regularization of solution ambiguity in underdetermined inverse and optimization problems, Journal of Computational Physics: X 13 (2022) 100105.
\newblock \href {https://doi.org/10.1016/j.jcpx.2022.100105} {\path{doi:10.1016/j.jcpx.2022.100105}}.

\bibitem{piroozmand23}
P.~Piroozmand, O.~Brenner, P.~Jenny, Dimensionality reduction for regularization of sparse data-driven {RANS} simulations, Journal of Computational Physics 492 (2023) 112404.
\newblock \href {https://doi.org/10.1016/j.jcp.2023.112404} {\path{doi:10.1016/j.jcp.2023.112404}}.

\bibitem{leoni20}
P.~C.~D. Leoni, A.~Mazzino, L.~Biferale, Synchronization to big data: nudging the {N}avier-{S}tokes equations for data assimilation of turbulent flows, Physical Review X 10~(1) (2020) 011023.
\newblock \href {https://doi.org/10.1103/physrevx.10.011023} {\path{doi:10.1103/physrevx.10.011023}}.

\bibitem{chen19}
J.~Chen, C.~Wolfe, Z.~Li, A.~Kyrillidis, Demon: improved neural network training with momentum decay (Oct. 2019).
\newblock \href {https://doi.org/10.48550/ARXIV.1910.04952} {\path{doi:10.48550/ARXIV.1910.04952}}.

\bibitem{fe50}
\href{https://sourceforge.net/p/foam-extend/foam-extend-5.0}{foam-extend-5.0} (2022).
\newline\urlprefix\url{https://sourceforge.net/p/foam-extend/foam-extend-5.0}

\bibitem{of8}
\href{https://openfoam.org}{{OpenFOAM} 8} (2020).
\newline\urlprefix\url{https://openfoam.org}

\bibitem{pontvilchez19}
A.~Pont-Vílchez, F.~X. Trias, A.~Gorobets, A.~Oliva, Direct numerical simulation of backward-facing step flow at and expansion ratio 2, Journal of Fluid Mechanics 863 (2019) 341--363.
\newblock \href {https://doi.org/10.1017/jfm.2018.1000} {\path{doi:10.1017/jfm.2018.1000}}.

\bibitem{gloerfelt19}
X.~Gloerfelt, P.~Cinnella, Large eddy simulation requirements for the flow over periodic hills, Flow, Turbulence and Combustion 103~(1) (2019) 55--91.
\newblock \href {https://doi.org/10.1007/s10494-018-0005-5} {\path{doi:10.1007/s10494-018-0005-5}}.

\bibitem{webster96}
D.~R. Webster, D.~B. Degraaff, J.~K. Eaton, Turbulence characteristics of a boundary layer over a two-dimensional bump, Journal of Fluid Mechanics 320 (1996) 53--69.
\newblock \href {https://doi.org/10.1017/s0022112096007458} {\path{doi:10.1017/s0022112096007458}}.

\bibitem{matai19}
R.~Matai, P.~A. Durbin, Zonal eddy viscosity models based on machine learning, Flow, Turbulence and Combustion (Feb. 2019).
\newblock \href {https://doi.org/10.1007/s10494-019-00011-5} {\path{doi:10.1007/s10494-019-00011-5}}.

\end{thebibliography}
\end{document}